\documentclass[preprint2]{aastex}

\slugcomment{For publication in AJ}

\begin{document}

\title{Star-Formation Rates of Local Blue Compact Dwarf Galaxies I: 1.4 GHz and 60\,$\mu$m luminosities}
\author{A. M. Hopkins\altaffilmark{1,3},
 R. E. Schulte-Ladbeck\altaffilmark{1}, I. O. Drozdovsky\altaffilmark{1,2}}

\altaffiltext{1}{Department of Physics and Astronomy, University of
  Pittsburgh, Pittsburgh, PA, 15260, USA}
\altaffiltext{2}{Astronomical Institute, St.Petersburg State University,
        Petrodvoretz, 198904, Russia}
\altaffiltext{3}{Hubble Fellow; ahopkins@phyast.pitt.edu}

\begin{abstract}
We determine and examine the star-formation rates of 50 well known,
local Blue Compact Dwarf (BCD) galaxies based on their 1.4\,GHz and
60\,$\mu$m luminosities. We find that in cases for which both radio
and far-infrared luminosities are available, the resulting star-formation
rates agree extremely well with one another. We determine that the
star-formation rates of the BCD galaxies in our sample span nearly five orders
of magnitude, from approximately a few times 10$^{-3}$ to several 
times 10$^{1}$\,M$_\odot$\,yr$^{-1}$, with a median SFR
of about 0.3$\,$M$_\odot$\,yr$^{-1}$. We discuss trends of metallicity
(primarily oxygen abundance) with star-formation rate, and explore
the connections between SFR and galaxy mass estimates.
\end{abstract}

\keywords{galaxies: dwarf --- galaxies: starburst --- galaxies: evolution}

\section{Introduction}

Blue Compact Dwarf (BCD) galaxies provide evidence that isolated, small,
low-metallicity galaxies may experience vigorous star formation in the present
epoch, challenging our current understanding of galaxy formation. The very
nature of BCDs remains puzzling in spite of thirty years of research since
they were first recognized as an interesting class of galaxy \citep{SS:70}.
For instance, it is still debated whether some ``primeval" BCDs are perhaps
in the process of undergoing their very first starburst, or whether
BCDs undergo several episodic starbursts separated by long quiescent
intervals, what triggers the star formation in BCDs, and whether
starburst-induced supernova-driven winds can cause the transformation
of BCDs into one of the other known types of dwarf galaxy
\citep[for a recent review]{KO:00}. While star-formation rates (SFRs)
are central to discussions of the evolution of BCDs, there are surprisingly
few quantitative SFRs for BCDs available in the literature.

The earliest and most quoted quantitative SFRs are based on direct measures
of the ultraviolet luminosity of massive stars in formation.
\citet{Fan:88} modeled the International Ultraviolet Explorer
spectra of seven BCDs. The implied SFRs for a Salpeter initial mass
function (IMF) with index 2.35 and mass cutoffs at 0.1 and 100\,M$_\odot$
(which we adopt for all measurements given in this paper) range from about 
$2.5\times10^{-3}$ to 25\,M$_\odot$\,yr$^{-1}$.

Subsequent efforts preferentially used the reprocessed ultraviolet radiation
of massive stars as measures for BCD SFRs, derived from either dust, or
ionized gas emission. \citet{Sage:92} investigated a sample of
15~BCDs, and derived SFRs based on far-infrared (FIR) and H$\alpha$ luminosity.
Their values range from about 0.1 to 30\,M$_\odot$\,yr$^{-1}$. \citet{Pop:99}
studied the SFRs of a sample of 48 compact star-forming dwarf galaxies in
voids. Using spectroscopically derived
H$_{\beta}$ luminosities, they determined values that range from
0.02 to 2.2\,M$_\odot$\,yr$^{-1}$, with a mean SFR of about
0.5\,M$_\odot$\,yr$^{-1}$. Most recently, \citet{Izo:00}
derived the SFRs of 27 BCDs from the Second Byurakan Survey (SBS). Their SFRs
are based on FIR (60\,$\mu$m) and radio continuum (1.4\,GHz)
luminosities. The SFRs derived from FIR luminosity span the range
from about 0.9 to 80\,M$_\odot$\,yr$^{-1}$. Radio luminosities were
converted to SFRs assuming that the continuum is due to thermal,
free-free emission. These SFRs range from approximately 2.5 to
270\,M$_\odot$\,yr$^{-1}$, with a mean of about 18\,M$_\odot$\,yr$^{-1}$.

The values available in the literature suggest that the SFRs of BCDs might
vary over five orders of magnitude, from 10$^{-3}$ to
10$^{2}$\,M$_\odot$\,yr$^{-1}$. However, the mean SFRs given in the existing
investigations are noted to also vary over two orders of magnitude, from
$0.1$\,M$_\odot$\,yr$^{-1}$ to 10\,M$_\odot$\,yr$^{-1}$. It is therefore
not clear whether and to what extent the wide range of BCD SFRs 
results from the use of different techniques, assumptions, or sample
selections, employed in the available studies. Clearly, a systematic
exploration of the parameter space of BCD SFRs is both warranted and
timely. Accurate SFRs are needed before any correlations of the SFR with
other physical properties can be discussed.

This paper is meant to be a first step toward establishing a baseline of
systematic, quantitative SFRs for BCDs. It was motivated by our
desire to measure the SFRs of BCDs using a uniform approach and by
employing techniques that are prone to minimal systematic errors.
Specifically, as star-formation may occur throughout BCDs and not just in
their very active centers, we excluded spectroscopic methods. These
require corrections for the (unknown) amount of flux missed outside of
the aperture. We also wanted to use a measurement that is free from 
the potentially large and uncertain corrections imposed by internal and
foregound dust extinctions. This immediately also ruled out ultraviolet
continuum and optical emission line indicators of the SFR. We therefore
decided to examine the SFRs of BCDs based on their radio continuum and
FIR luminosities. Fortunately, public databases of FIR and radio flux
densities that include BCDs are readily at hand.
For completeness, and to investigate the extent of and role played by
obscuration in our sample, a subsequent study to be presented in Paper~II
will explore the SFRs derived from very wide-slit H$\alpha$ spectroscopy.

\section{Sample selection}

The sample under investigation consists of 50 of the most well known and
well studied BCDs. It is in this regard that our sample consists of
``typical" BCDs, and that the resulting SFRs are called ``typical".
By no means does this investigation present an unbiased or a complete
sample of BCD SFRs. The present sample comes from those BCDs that have
been well studied because they were used to derive the primordial helium
abundance, and have been taken from a recent compilation of 54 objects
\citep[hereafter IT]{IT:99}. For all of these galaxies, accurate
redshifts, needed to calculate luminosities, are available. The BCDs in
this sample also have careful determinations of the oxygen abundance of
the ionized gas. Since oxygen is produced by massive stars, and since
massive stars are used as a quantitative measure of the SFR, a particular
emphasis of the present paper is to investigate any correlations between
these two quantities.

Observational information for the galaxies in this study is presented in
Table~\ref{tab1}. This contains positions (J2000), $60\,\mu$m flux densities
from IRAS (Infrared Astronomical Satellite) catalogues, $1.4\,$GHz flux
densities from the NVSS (NRAO VLA Sky Survey) and FIRST (Faint Images
of the Radio Sky at Twenty centimeters) catalogues and images, and
heliocentric velocities. Much of this information was obtained from
NED\footnote{The NASA/IPAC Extragalactic Database (NED) is operated by the
Jet Propulsion Laboratory, California Institute of Technology,
under contract with the National Aeronautics and Space Administration.}.
Four of the galaxies in Table~\ref{tab1}, Mrk~71, SBS~1319+579, SBS~1331+493
and SBS~1533+574, were split into multiple components for their metallicity
measurements \citep[their Table~1, which has 54 entries]{IT:99}.
Since the $60\,\mu$m (IRAS) and 1.4\,GHz (NVSS) flux densities do not have
fine enough resolution to distinguish between these components, these
galaxies are treated as single objects in the current study. The
metallicities (shown in Table~\ref{tab2}) for those objects are the average
of the values measured for the individual components (IT), while the
errors reflect the ranges in values.

\subsection{Distances and flux densities}

\begin{figure*}
\centerline{\includegraphics[width=15.0cm]{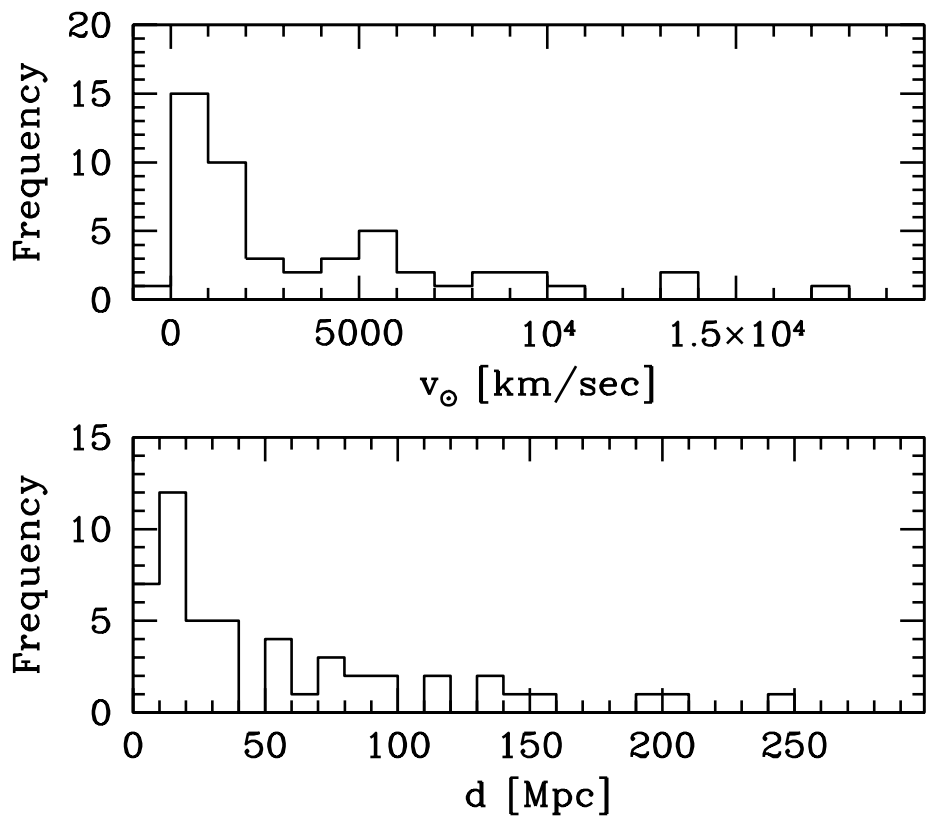}}
 \caption{Histograms showing the heliocentric velocities and
Virgo-corrected distances for the sample.
 \label{disthist}}
\end{figure*}

The majority of the 50 BCDs in the sample had their distances estimated
based on heliocentric velocities taken from NED. The velocities given
in NED were derived from emission line or HI measurements made predominantly
by Izotov, Thuan and collaborators \citep{Thu:99a,IT:98,Izo:97,Izo:94}, or
taken from the RC3 \citep{deV:91}. In cases where velocities for several
components were listed, we assumed the velocity that is associated with
the main body or main component of the galaxy. Distances for all but three
of our sample were estimated from velocities by using a Virgocentric flow
model \citep{Kra:86}. We adopted a Hubble constant of
$72\pm 8$\,km\,s$^{-1}$\,Mpc$^{-1}$ \citep{Fre:01} and a Virgo distance
of $17.0\pm 0.3$\,Mpc \citep{Ton:01}. The three exceptions, where
distances were estimated from analysis of the I-band brightness of tip of
the red giant branch (TRGB) stars in color-magnitude diagrams (CMDs) of
stellar populations, are UGC~04483 \citep{Dol:01}, VII~Zw~403 \citep{Sch:99}
and Mrk~0209 (I~Zw~36) \citep{Sch:01a}. We present histograms showing the
distributions of distances and heliocentric velocities for our BCD sample
in Figure~\ref{disthist}.

The ratios of the distances from the Virgocentric flow model to the TRGB
distances for UGC~04483, VII~Zw~403 and Mrk~0209 are 1.69, 0.23, and $<$1.2,
respectively. This indicates a considerable misjudgment of the distances
based on velocities and the Virgocentric flow model is possible. We believe
that distance errors can potentially amount to up to a factor of a few, but
probably do not exceed an order of magnitude. It is this difficulty in
assessing distance errors that led us to omit errors on the SFRs derived
below. 

\subsubsection{IRAS}

The Infrared Astronomical Satellite has proved an immensely valuable resource
in exploring the infrared properties of galaxies since its launch in 1983.
IRAS products include 60\,$\mu$m flux densities for
galaxies from the Point Source Catalog (brighter than about 0.5\,Jy) and
Faint Source Catalog (typically brighter than 0.2\,Jy). These fluxes were
obtained for the BCD sample, where available, from NED and from the
IRSA\footnote{The NASA/IPAC Infrared Science Archive 
is operated by the Jet Propulsion Laboratory, California Institute of
Technology, under contract with the National Aeronautics and Space
Administration. See http://irsa.ipac.caltech.edu/} ISSA image server
and catalog overlay service. The resolution of the IRAS images at
60\,$\mu$m is about $1'$ with typical positional accuracies better than
$20''$. Of the 50 BCDs in our sample 18 are detected at 60\,$\mu$m by IRAS.
We have not made direct measurements for the flux density upper limits of
the non-detections, but for most of these the 0.2\,Jy limit of the Faint
Source Catalog should be indicative.

\subsubsection{NVSS}

The NRAO VLA Sky Survey \citep{Con:98} is a 1.4\,GHz radio continuum
survey covering the sky north of $-40^{\circ}$ declination,
made using the Very Large Array (VLA) operated by the National
Radio Astronomy Observatory (NRAO). The observations were made
using the compact D and DnC configurations of the VLA. The
resulting images have an rms noise level of about 0.5\,mJy\,beam$^{-1}$
and a FWHM resolution of $45''$. Positional accuracies (from
the source measurements using a gaussian fitting algorithm) range
from about $1''$ for sources stronger than 15\,mJy to $7''$
at the survey limit. The catalog lists sources detected above a peak
brightness of 2\,mJy\,beam$^{-1}$. For 13 of the galaxies in our BCD sample
an NVSS catalog source is close enough ($< 30''$) to be a probable
identification, and the flux measurement from the catalog
was taken. For another 10 of the sample galaxies, where the NVSS
FITS image showed a local peak in intensity at the position of the galaxy,
but no catalog source was present, we performed our own source measurements.
We used the {\sc miriad} (Multichannel Image Reconstruction, Image
Analysis and Display) task {\sc sfind} \citep{Hop:02} to make these
measurements, and the sources so detected are typically at the
$4-5\,\sigma$ level. For non-detections, we measured the local rms
noise level in the vicinity of the BCD galaxy and adopted 5 times
this value as the upper limit on the NVSS flux density.
Uncertainties in the flux density measurements of the NVSS catalogue
have been described in detail \citep[see also their Figure~31]{Con:98}.
In practice, these uncertainties rise from $\approx 5\%$ at flux
densities of 10\,mJy to $\approx 30\%$ close to the survey limit.
These uncertainties translate to uncertainties in estimated
star-formation rates (see Section~\ref{SFR}).
Digitized Sky Survey\footnote{The Digitized Sky Surveys
were produced at the Space Telescope Science Institute under U.S. Government
grant NAG W-2166. The images of these surveys are based on photographic data
obtained using the Oschin Schmidt Telescope on Palomar Mountain and the
UK Schmidt Telescope. The plates were processed into the present
compressed digital form with the permission of these institutions.}
images have been overlayed with 1.4\,GHz contours
from NVSS for each of the 23 detected BCDs (Figure~\ref{nvss1}).

\subsubsection{FIRST}

FIRST, Faint Images of the Radio Sky at Twenty centimeters
\citep{Bec:95}, is a 1.4\,GHz radio continuum survey currently
covering about 8500 square degrees of the North Galactic Cap.
The observations were taken using the VLA in B configuration.
The images have a typical rms noise level of 0.15\,mJy with
a FWHM resolution of $5''$. Source positional accuracies 
vary from $0.5''$ at the 3\,mJy level to about $1''$ for the
faintest sources. The source detection threshold of the survey
is 1\,mJy. Because of the much higher resolution of FIRST
compared to NVSS, some unresolved NVSS objects are resolved
by FIRST into more than one component (UM~311 is one example).
In this case we have taken the sum of the FIRST flux densities
for the individual components as the FIRST measurement
for the whole source. From the FIRST catalog, 6 galaxies were
matched with our BCD sample, and we obtained upper limits from
the FITS images for another 14. The flux density upper limits
are again taken to be 5 times the local rms value. The remaining
30 BCDs lie outside the area of the FIRST survey. 1.4\,GHz contours
from FIRST are shown on the images in Figure~\ref{nvss1} for each
of the 6 detected BCDs.

An important advantage from the higher resolution of FIRST images
is being able to identify the positional origin of the radio emission
with greater accuracy than allowed by NVSS. The different resolutions
are similarly an important effect to account for if calculating SFR per
unit area, and also allow for distinguishing between the object of
interest and confusing background sources. In the case of Mrk~0209,
for example, the FIRST emission appears to originate slightly
to the SE of the region of brightest optical emission. This may
initially suggest a background source for the radio emission, but the
strong consistency between the radio-derived SFR and numerous other
optically-derived values for this galaxy, \citep[see also discussion
in Section~\protect\ref{fircor}]{Sch:01a}, argue that in fact this
radio source is likely to be a star-forming knot in the outskirts of Mrk~0209.
Similarly, CG~0798 shows a double-peak in the NVSS emission, and
a FIRST source about an arcminute to the south of the galaxy accounts
for some of this emission. Lacking more detailed observations, it
is certainly possible that this may be a coincidental background source
and the radio flux-density and the derived SFR for CG~0798 should strictly
be treated as upper limits. Given the good agreement between the radio
and FIR SFRs for this galaxy, however, we have chosen to treat it as a
detection for the current analysis.

\section{Star-formation rates}
\label{SFR}

To avoid the effects of both spectroscopic flux losses and uncertain
obscuration corrections, we use 1.4\,GHz and 60\,$\mu$m luminosities
as estimators of the star-formation rate for galaxies in our BCD sample.
Here we adopt the calibrations
\begin{equation}
\label{eq1}
{\rm SFR_{60\mu{\rm m}}}=5.5\times\frac{L_{60\mu{\rm m}}}{5.1\times10^{23} ({\rm
 W\,Hz^{-1})}}
\end{equation}
\citep{Cram:98,Con:92} for the SFR derived from the $60\,\mu$m luminosity,
and
\begin{equation}
\label{eq2}
{\rm SFR_{1.4GHz}}=5.5\times\frac{L_{\rm 1.4GHz}}{4.6\times10^{21} ({\rm W\,Hz
^{-1})}}
\end{equation}
\citep{Haa:00,Con:92} for the SFR from the 1.4\,GHz luminosity.
In both cases the factor of 5.5 converts from the mass range
$5-100\,$M$_\odot$ to the mass range $0.1-100\,$M$_\odot$, assuming a
Salpeter IMF \citep[for example]{Haa:00}. Also, \citet{Haa:00} point out
that starburst galaxies may have IMFs that differ from the one adopted in
Equation~\ref{eq2}. In the worst case a maximum decrease of the SFR by a
factor around 5 is possible. Any changes to the slope of the IMF, however,
should affect the radio and FIR luminosities equally. 
The SFRs calculated for our BCD sample are given in Table~\ref{tab2}
along with the distances to the galaxies. The metallicities
(oxygen abundance) also shown have been taken from IT. HI masses have
been derived using HI flux densities from the literature with the
distances given here, and are also shown in Table~\ref{tab2}.

Distance uncertainties (possibly up to a factor of a few) are the dominant 
source of measurement uncertainty for the SFRs. For this reason, the
Figures below do not include error bars on the SFRs. Histograms showing the
distribution of the derived SFRs for our sample are presented in
Figure~\ref{sfrhist}. We also do not calculate SFR normalized to
the optical area of the galaxy. This is firstly because of the complex
optical morphologies for the objects in our sample (see Figure~\ref{nvss1}).
Secondly, the galaxies in our sample are essentially point sources for
the $60\,\mu$m and 1.4\,GHz detections. Instead, we investigate below the
properties not only of the total SFR, but also of the ``specific SFR" by
which we mean the SFR normalized by HI mass. This is a distance-independent
quantity, and a probe of relative star formation that should be reasonably
independent of galaxy size or mass. Details of the specific SFRs and HI masses
are given in Section~\ref{results} below.

The above calibrations of SFR from luminosity are derived independently
of metallicity, although it is worth noting that the 1.4\,GHz calibration
is based on the observed Galactic relation between supernova rate and
nonthermal radio luminosity. An implicit assumption in the case of
estimating SFR from the 1.4\,GHz luminosity is that there is no contribution
from an active nucleus (Seyfert or LINER activity) to the radio emission.
This assumption seems to be justified for the sample of BCDs we are
investigating, since all of them have HII-like spectra.
 
\section{Results}
\label{results}

The NVSS catalog yields flux densities for a significant fraction of our
sample. Table~\ref{tab2} and Figure~\ref{sfrhist} show that the SFRs derived
from NVSS fluxes cover almost five orders of magnitude. Clearly, there are
real differences in BCD SFRs, since the range seen is much larger than
the potential distance uncertainties. From the 23 NVSS-detected objects
we calculate a median SFR of 0.27$\,$M$_\odot$\,yr$^{-1}$. This compares
well with the median SFR, 0.36$\,$M$_\odot$\,yr$^{-1}$, of the 18
IRAS-detected BCDs. The IT sample of BCDs being examined, however, is
likely closer to being a flux-limited than a volume-limited sample,
favoring more luminous galaxies with higher SFRs, and the NVSS sensitivity
limit strengthens this bias. Given that the radio and FIR detections
are subject to such Malmquist bias, artificially raising the median value,
the median SFR quoted above should not be over-interpreted. To provide
some perspective, an approximate uncertainty for this median can be
estimated by examining the range of SFRs spanned by the
$3\,(\approx\sqrt{11})$ galaxies either side of Mrk~1416 (having the median
SFR$=0.27$). This gives a range of an order of magnitude, from
0.13 to 1.3$\,$M$_\odot$\,yr$^{-1}$.

Figure~\ref{sfrvssfr} shows the comparison between SFRs derived from 1.4\,GHz
(FIRST and NVSS) and 60\,$\mu$m luminosities. A good agreement exists
between the NVSS-based 1.4\,GHz SFRs and the IRAS-based 60\,$\mu$m SFRs.
It is evident from Table~\ref{tab1} and this Figure that the FIRST
flux densities tend to be systematically lower than those from the NVSS
for most of the galaxies detected by both surveys. This is also true for
many of the FIRST upper limits where an NVSS detection exists.
The probable reason for this is the difference between the two surveys
in their sensitivity to extended emission. FIRST, made using the
VLAs B-array, has much higher angular resolution than NVSS, made
with the more compact D and DnC arrays, but at the expense of
sensitivity to more extended emission. FIRST images are typically only
sensitive to structure extended over less than about $2'$, while the NVSS
images will detect emission from sources up to about $15'$ across
\citep{Per:00}. As a result the NVSS flux density measurements tend
to be somewhat larger than those from FIRST, since many of the galaxies
in our sample are larger than the extent over which FIRST is most sensitive.
In subsequent analysis where 1.4\,GHz-derived SFRs are used we thus rely
on SFRs derived from NVSS measurements, both for this reason and also
since more of our sample galaxies have NVSS detections.

Figure~\ref{sfrvsdist} presents SFR$_{\rm 1.4GHz}$ as a function of
distance. The NVSS detection threshold is marked. As may be expected from
the tight relation in Figure~\ref{sfrvssfr}, a diagram showing
SFR$_{60\mu{\rm m}}$ versus distance is very similar. An interesting
point to note from this Figure is the absence of any moderate
to high SFR objects closer than $\approx 10$\,Mpc, that is, in
the region where Hubble Space Telescope (HST) observations (with 
WFPC2 or NICMOS instruments) are judged able to resolve individual stars
within the target galaxies. This is mainly a result of high SFR objects
having lower space-densities than low SFR galaxies, combined with the
small volume being probed at this distance. The implication here is
that HST-based CMD investigations of nearby galaxies are restricted to
lower-SFR objects, in good agreement with the low SFRs derived
from CMD modeling.

The 1.4\,GHz flux on which SFR$_{\rm 1.4GHz}$ is based is thought to
measure the present-day SFR on a similar timescale as the lifetime of
massive stars \citep{Con:92}. We now explore the correlation between
various chemical abundances from IT and absolute and specific SFR.
The chemical elements that we emphasize are associated with progenitors of
different masses, and this translates into different enrichment timescales.
Specifically, oxygen is produced by massive stars and returned quickly to
the interstellar medium (ISM) by way of supernova (SN) type II explosions, 
nitrogen is thought to be produced in intermediate-mass stars and returned
to the ISM in the asymptotic giant branch (AGB) wind phase (but see IT for
a contrary view), while iron is considered to reflect the ISM's pollution
by low mass stars through SN\,Ia events, and should lag the most behind the
oxygen enrichment. 

In the top panel of Figure~\ref{zvssfr} a weak positive trend of
12+log(O/H) (from IT) with absolute SFR$_{\rm 1.4GHz}$ can be seen. The
Spearman rank test shows that the null hypothesis (no correlation between
12+log(O/H) and log(SFR)), using only the detections, has a probability of
only about 0.6\%. An unweighted linear regression yields
12+log(O/H) = 0.09log(SFR) + 7.90 (with the uncertainty of the slope being
$\sigma_a$ = 0.04 and the uncertainty of the intercept being
$\sigma_b$ = 0.05). The extremely metal-poor galaxies I~Zw~18 and 
SBS~0335$-$052 lie well away from the trend shown by the majority of the
sample, effectively lowering the intercept. If we exclude I~Zw~18 and
SBS~0335$-$052 from the sample, than the Spearman rank correlation coefficient
increases slightly, and the probability of a chance correlation drops to just
below 0.5\%, while the linear regression gives
12+log(O/H) = 0.09log(SFR) + 7.96 (with $\sigma_a$ = 0.03
and $\sigma_b$ = 0.03). It is this linear fit that is shown in
Figure~\ref{zvssfr} as a solid line, with the dashed lines reflecting the
error in the intercept. (Notice the fit results reported here are slightly
different from the preliminary results of \citet{Sch:01b}
because here we use more up-to-date values for the Hubble constant and the
Virgo distance.) We conclude that BCDs with high present-day total
SFRs clearly also have high oxygen abundances.

If we replace absolute SFRs with specific SFRs (i.e.\ SFR/M$_{\rm HI}$)
a weak positive trend seems to remain (HI masses are discussed below). The
Spearman rank test, however, indicates the hypothesis that 12+log(O/H) and
log(SFR/M$_{\rm HI}$) are correlated has a 15\% probability of occuring
by chance. As a result it seems there is little correlation, if any,
between oxygen abundance and SFR/M$_{\rm HI}$.

In the middle panel of Figure~\ref{zvssfr} a weak positive trend of
log(N/O) (from IT) with absolute SFR can be seen. The extremely metal-poor
galaxies I~Zw~18 and SBS~0335$-$052 do not distinguish themselves on this
plot, lying well within the distribution of values. The Spearman rank test
shows that the probability for a chance correlation is about 0.7\%. A linear
regression yields log(N/O) = 0.09log(SFR) $-$ 1.42
(with $\sigma_a$ = 0.02 and $\sigma_b$ = 0.03). Conversely, the Spearman rank
test for log(N/O) with the logarithm of the specific SFR,
log(SFR/M$_{\rm HI}$), shows the probability for a chance correlation is 5\%.

Lastly, we investigate the dependence bewteen [O/Fe] (from IT) and log(SFR)
or log(SFR/M$_{\rm HI}$) (see bottom panel of Figure~\ref{zvssfr}).
The Spearman rank test gives such small correlation coefficients that the
data are likely to be uncorrelated. The linear regression of [O/Fe] on
log(SFR) results in a constant, [O/Fe] = 0.43
(with $\sigma_a$ = 0.04 and $\sigma_b$ = 0.04). Figure~\ref{zvssfr} shows 
that SBS~0335$-$052 is quite removed from the rest of the sample BCDs, but
I~Zw~18 lies well within the distribution. 

Figure~\ref{sfrvsw20} explores the dependence of various quantities on galaxy
mass. The width of the HI emission line profile at 20\% of peak, W$_{20}$,
is taken as an indicator of galaxy total mass. HI profile widths are often
used to calculate galaxy total masses, but require radii and inclinations.
As can be seen from Figure~\ref{nvss1}, radii and inclinations are difficult
to judge for BCDs because the star-forming regions dominate their appearance.
W$_{20}$, on the other hand, also may not be a solid indicator for
rotational velocity if turbulent motions are significant.

With these caveats in mind, we plot in the upper panel of Figure~\ref{sfrvsw20}
the HI mass versus the W$_{20}$ velocity. HI and W$_{20}$ data were taken
from \citet{Thu:99a} (circles) or \citet{TM:81} (triangles). 
The Spearman rank correlation coefficient of log(M$_{\rm HI}$) and
log(W$_{20}$) is high, indicating a less than 0.05\% probability that
this is a chance correlation. In other words, we can be confident
that HI mass increases as total mass increases. A linear regression gives
log(M$_{\rm HI}$) = 3.6log(W$_{20}$) + 1.3
(with $\sigma_a$ = 0.6 and $\sigma_b$ = 1.3), and is shown in
Figure~\ref{sfrvsw20}.

The Spearman rank test for total SFR and total mass rejects the null
hypothesis; correlation by chance has about a 0.5\% probability (middle
panel of Figure~\ref{sfrvsw20}). The linear regression is
log(SFR) = 4.8log(W$_{20}$) $-$ 10.5 
(with $\sigma_a$ = 1.0 and $\sigma_b$ = 2.0). The higher SFRs are clearly
being hosted by higher mass galaxies, although there is a significant
scatter in the trend (for a given rotational velocity or total mass,
the SFR may vary over more than an order of magnitude). 

Finally, we also investigate the correlation between metallicity and
total mass. The lower panel of Figure~\ref{sfrvsw20} shows that the
lowest metallicity objects, such as I~Zw~18 or SBS~0335$-$052, do not
have the lowest rotational velocities. Indeed the Spearman rank test
indicates that the probability of a chance correlation between 12+log(O/H)
and log(W$_{20}$) is quite high, almost 10\%. A higher total mass does
not necessitate a higher oxygen abundance.

Figure~\ref{sfrvsmhi} shows the variation of SFR$_{\rm 1.4GHz}$ with
HI mass. The data were taken from \citet{Thu:99a} (circles) 
and from \citet{TM:81} (triangles). Because of our interest
in extremely low-metallicity objects, and because it is one of the
BCDs with a TRGB distance, we added UGC~04483 to the Figure, using the
HI flux from \citet{HR:86}. This data point is shown
as a square. The higher SFRs are clearly being hosted by galaxies with
higher HI masses although there is a significant scatter in the trend
(again, as with W$_{20}$, for a given HI mass the SFR may vary
over more than an order of magnitude). This scatter is reflected in
the wide range of timescales for gas consumption, from 0.3\,Gyr
to more than 10\,Gyr, indicated by the solid lines in Figure~\ref{sfrvsmhi}.
I~Zw~18 and SBS~0335$-$052 do not stand out in any way from the rest of
the sample in this picture.

The results can be summarized as follows:
\begin{enumerate}
\item The median SFR in the IT sample is 0.3$\,$M$_\odot$\,yr$^{-1}$
(with an uncertainty up to a factor of three or so, and it may be an upper
limit due to Malmquist bias).
\item Total SFRs in the IT sample range over at least four orders of magnitude.
\item SFRs calculated from IRAS and from NVSS data are consistent with
SFR$_{60\mu{\rm m}}$ = SFR$_{\rm 1.4GHz}$, i.e., these galaxies obey
the FIR/radio correlation valid for large spirals.
\item While BCDs in this sample are close, z$<$0.06, there is still a
lack of high-SFR galaxies within the range of HST single-star photometry
(about 10\,Mpc). This is because such objects have space densities low
enough that they are unlikely to appear in the small volume probed out
to this distance.
\item We find a high probability of a positive correlation between
12 + log(O/H) or log(N/O) and total SFR, but [O/Fe] and total SFR appear
to be uncorrelated. For all three element abundances, the probability of
a correlation with SFR/M$_{\rm HI}$ (a distance-independent quantity) is low.
\item Taking W$_{20}$ as an indicator of total mass, we find a high
probability for a positive correlation between HI mass and total mass,
and between total SFR and total mass. The probability for a correlation
of metallicity and total mass, on the other hand, is small.
\item The ``typical" timescale for gas consumption in the IT BCD sample
is a few Gyrs. None of these galaxies is in danger of running out of gas
in the very near future. However, these BCDs will die sooner than the
typical dwarf Irregular galaxy.
\end{enumerate}

\section{Discussion}

\subsection{The Far-Infrared/Radio Correlation for BCDs}
\label{fircor}

We find excellent agreement between SFR$_{60\mu{\rm m}}$ and
SFR$_{\rm 1.4GHz}$ using the calibration equations of \citet{Con:92},
\citet{Cram:98}, and \citet{Haa:00}. Specifically, and as
shown by Figure~\ref{sfrvssfr}, the data are consistent with
SFR$_{60\mu{\rm m}}$ = SFR$_{\rm 1.4GHz}$. This consistency, however,
does not necessarily imply that the correct SFR has been determined.
As an independent check of the SFRs from FIR and radio luminosities we
compare them with results from modeling of color-magnitude diagrams.
CMD studies have their own but different set of limitations, in terms
of sensitivity, color biasing, and fractional galaxy area observed.
We have only two cases in the IT sample for which there are both distances
and present-day SFRs available from single-star photometry. For VII~Zw~403,
SFR$_{60\mu{\rm m}}=0.0096\,$M$_\odot$\,yr$^{-1}$ and
SFR$_{\rm 1.4GHz}=0.004\,$M$_\odot$\,yr$^{-1}$, while CMD modeling of
HST/WFPC2 V and I photometry yields 0.011$\pm$0.008$\,$M$_\odot$\,yr$^{-1}$
\citep{Cro:02}. For Mrk~209 (I~Zw~36) the
SFR$_{\rm 1.4GHz}$ is $0.0217\,$M$_\odot$\,yr$^{-1}$, whereas HST/NICMOS J
and H data were modeled to yield a SFR of
$0.025\pm0.004\,$M$_\odot$\,yr$^{-1}$ \citep{Sch:01a}. We conclude
that CMD modeling validates the FIR and radio SFRs to the extent
that it is possible with the limited amount of data.

Astrophysical interpretations of the FIR/radio correlation suppose
that both types of emission have their origin in newly forming massive stars.
The FIR emission originates from the heating of dust by
UV radiation, while the radio emission is produced by synchrotron radiation
from cosmic ray electrons accelerated by shocks associated with supernova
remnants \citep{Xu:90}. Both SFR$_{60\mu{\rm m}}$ and SFR$_{\rm 1.4GHz}$
are considered to be
measures of the present-day SFR. The synchrotron component of the radio
emission is thought to dominate for a star-forming galaxy observed at
1.4\,GHz, with a much smaller contribution from the thermal Bremsstrahlung
of the HII regions. With a typical synchrotron spectrum for the nonthermal
component ($S_{\nu}\propto\nu^{-0.8}$), less than $1/8$ of the total
1.4\,GHz flux is thermal \citep{Con:92}. Equation~\ref{eq2} above makes
the assumption that supernovae generate similar amounts of radio luminosity
to that observed in our own Galaxy, and the detection of any nonthermal
radio luminosity in external star-forming galaxies implies the presence of
magnetic fields.

The calibration above for SFR$_{\rm 1.4GHz}$ differs from the calibration
given by \citet{Izo:00}. \citeauthor{Izo:00} studied SFRs in a sample of
BCDs drawn from the Second Byurakan survey, but they find
SFR$_{\rm 1.4GHz} \approx 3.5$~SFR$_{60\mu{\rm m}}$. Their
calibration for SFR$_{\rm 1.4GHz}$ from 1.4\,GHz luminosity makes the
assumption that the radio emission is purely thermal, although they
acknowledge that a significant part of the 1.4\,GHz radio emission should
be nonthermal emission related to supernova events. This explains their
overestimation of 1.4\,GHz luminosity derived SFRs. The calibration for
SFR$_{60\mu{\rm m}}$ given by \citeauthor{Izo:00} also differs slightly from
the one adopted above (but see discussion of SFR$_{\rm FIR}$ calibrations
given by \citeauthor{Ken:98} \citeyear{Ken:98}). It can easily be shown
that the objects in the sample of \citeauthor{Izo:00} follow the FIR/radio
correlation, and indeed the SFRs derived for their sample by using the
calibrations of Equations~\ref{eq1} and \ref{eq2} result in highly consistent
SFRs from both $60\,\mu$m and 1.4\,GHz luminosities.

There are suggestions that the thermal component of the radio luminosity
in BCDs is larger than that typical for normal spirals. \citet{Kle:91}
found the radio spectra of BCDs are generally flatter than those of spiral
galaxies. They comment that they find it surprising that BCDs do follow
the same FIR/radio correlation as do normal spirals, given the strongly
varying SFRs in BCDs (further discussed below) and the lack of cosmic-ray
confinement. They suggest that a lower ratio of nonthermal-to-thermal
radio emission in BCDs could be offset by a similar deficiency in
FIR emission. \citet{Dee:93} re-observed a few of these BCDs 
and also found a variety of shapes for their radio spectra, but conclude
that a nonthermal component is present in most. Clearly, these data
suggest that the radio spectra of BCDs at 1.4\,GHz have differing amounts
of nonthermal-to-thermal emission, hence it is indeed surprising that
our investigation yields identical FIR and 1.4\,GHz SFRs when the
calibrations of Equations~\ref{eq1} and \ref{eq2} are applied.
There are no linear radio polarization observations of BCDs in the
literature, and therefore we have no direct measurements of their
large-scale magnetic fields. The presence of a one-to-one FIR-radio
correlation in our BCD sample suggests that magnetic fields in these
systems need to be of at least sufficient strength to extract synchrotron
radiation from relativistic electrons accelerated by supernova remnants.

The conversion of ${60\mu{\rm m}}$ luminosity to SFR$_{60\mu{\rm m}}$ rests
on the assumption that dust absorbs well in the UV, and that it re-radiates
effectively in the IR, so that the ultraviolet radiation of massive, young
stars is absorbed and re-emitted as FIR luminosity. Clearly the
conversion factor must depend on the amount of dust present. There has been
a persistent notion in the literature that metal-poor dwarf galaxies are
virtually dust free. This is not the case. We emphasize that although our
Table~\ref{tab2} shows no FIR detections for I~Zw~18 and SBS~0335$-$052,
it is now known that both contain appreciable amounts of dust
\citep{Thu:99b,Dal:01,Can:01}. In any event, for the
SFR$_{60\mu{\rm m}}$/SFR$_{\rm 1.4GHz}$ ratio to be unity, it is not as
much the total amount of dust in dwarfs that matters, but rather, that
there is sufficient dust to produce a FIR luminosity that offsets the
radio luminosity. This idea supports the suggestion that the local properties
of individual star-forming regions are more important in generating the
FIR/radio correlation than the global properties of the galaxies that
host them.

Is metallicity likely to affect the calibrations of SFR to luminosity?
\citet{LF:98} discuss dust-to-gas ratios as a function of
metallicity in dwarf galaxies. \citet{Hir:01} incorporate
metallicity into the conversion factor, and find it can vary by about
a factor of 5 through differences in dust content (dust-to-gas ratios)
between our Galaxy and others. In order to investigate whether there is
any metallicity bias in the SFRs we derived for BCDs, we display in
Figure~\ref{sfrratio} SFR$_{\rm 1.4GHz}$/SFR$_{60\mu{\rm m}}$
versus 12+log(O/H). This ratio spans a range of about 0.3 to 1.8
for metallicities of 12+log(O/H) = 7.7 to 8.3. As can be seen from
this Figure we do not have measurements at both frequencies for a number
of the extremely metal-poor objects. For three objects with a $60\,\mu$m
detection but a 1.4\,GHz upper limit, we see upper limits to this ratio
between 0.8 and 1.5. For six objects with 1.4\,GHz detections but $60\,\mu$m
upper limits, this ratio shows lower limits ranging from 0.6 to 2.5. Both
I~Zw~18 and SBS~0335$-$052 fall into the latter category and have lower
limits of 1.5. The mean metallicity of the detected sample is
12+log(O/H) = 7.98. IT use the scale on which the Sun has
12+log(O/H) = 8.91, so we are effectively exploring the FIR/radio correlation
at about 0.1 of the solar metallicity. In our sample, the mean
SFR$_{\rm 1.4GHz}$/SFR$_{60\mu{\rm m}}$ is 1.02$\pm$0.40, identical to
the expected value of 1. No trend with metallicity is seen.

In starburst galaxies with highly episodic star-formation histories (SFH),
a time-delay between the maximum of the FIR and the nonthermal radio
emission has been hypothesized \citep{Lis:97}. Thermal radiation and
re-radiation respond instantaneously to the presence of newly born massive
stars, whereas nonthermal radiation does not occur until the end of the life
of massive stars, when they explode as SNe, some 10$^7$~yrs later.
If time-delays were important we would expect a large scatter of
SFR$_{\rm 1.4GHz}$/SFR$_{60\mu{\rm m}}$ about the mean, depending
on exactly when we observe a galaxy in its changing SFH. Indeed,
BCDs might be suspected of this effect if they experience highly episodic
starbursts, whereby very short bursts \citep[of about 3.5~Myr,][]{MK:99}
are separated by very long quiescence periods of a Gyr or more
\citep[but see][]{Sch:01a}. \citet{WF:01} recently modeled the size of
the offset in the H$\alpha$ luminosity to SFR$_{\rm H\alpha}$ calibration
that would arise due to episodic starbursts. The conversion factor ranges
over two orders of magnitude for the benchmark SFHs that they explore.
We observe a 1~$\sigma$ variation of $\pm$0.40 about the mean ratio for
SFR$_{\rm 1.4GHz}$/SFR$_{60\mu{\rm m}}$, a much smaller scatter, although
we admittedly are investigating an incomplete sample. It is possible that
time-delay effects contribute to this scatter about the mean,
but they clearly do not dominate the conversion factor.

To summarize, our finding of a metallicity-independent FIR/radio ratio of
unity implies that the conversion factors that are currently being used
to turn FIR and radio luminosity to SFR seem to remain valid at 0.1 of
the solar metallicity. (In Paper~II, we shall explore the conversion factor
that translates H$\alpha$ luminosity to SFR$_{\rm H\alpha}$ for these
objects.) We also infer that BCDs are required to have supernova rates,
radio luminosities per SN, magnetic fields, and dust contents in some
scaled proportion to those of the large spirals, to which the FIR and
radio calibrations were originally tailored.

\subsection{SFR, Metallicity, and Mass}

A major difficulty for chemical evolution models of BCDs has been to
explain their low oxygen abundances in spite of their active star formation.
In this section we confront the data assembled here with current theoretical
models for dwarf galaxy evolution.

We see a correlation of HI gas mass with total mass, and between
total SFR and total mass, in the sense that galaxies with more total mass
also have a higher gas mass, and support a higher present-day SFR. This comes
as no surprise since the large range of SFRs among different galaxies is
attributed in part to the large range in galaxy masses \citep{Ken:98}.
We see a broad trend of increasing SFR with increasing HI mass, 
but there is also a wide range of SFRs for a given HI mass. This is
explained as being due to different gas-consumption timescales. Here we have
a hint of some variation in the physical properties of BCDs. The gas depletion
timescale of BCDs in the IT sample is typically a few Gyrs. This is an order
of magnitude shorter than those in the dwarf Irregular sample studied by
\citet{vZ:01}. Our data also show a positive correlation between oxygen
abundance and SFR, in the sense that galaxies with higher global SFRs
also show higher oxygen abundance. This similarly comes as no surprise as the
massive stars that are used to gauge the SFR are also the sites of oxygen
nucleosynthesis. (However, there is some as yet unsettled discussion in
the literature as to whether the oxygen so measured is related to the
present generation of stars, or whether it stems from prior ones, e.g.\
\citeauthor{Lar:01} \citeyear{Lar:01} and references therein.)

While SFR correlates with total mass and oxygen abundance correlates
with SFR, there is {\em no correlation between oxygen abundance and
total mass\/} in our sample (see Figure~\ref{sfrvsw20}). We further find that
SFR/M$_{\rm HI}$ versus HI mass shows little correlation, if any, and we
also find no correlation between oxygen abundance and SFR/M$_{\rm HI}$.
\citet{Leq:79} first presented a relationship between metallicity
and total mass for a small sample of Irregular galaxies and two BCDs
(I~Zw~18 and II~Zw~40). \citet{Ski:89} later derived the well known
metallicity-luminosity relation for Irregular galaxies. In Irregular galaxies,
the oxygen abundance rises with absolute blue magnitude of the host galaxy
\citep[but see][for the opposing view]{HO:98,HH:99}. Note that \citet{Ski:89}
excluded BCDs from their sample. The metallicity-luminosity relation is
frequently interpreted as a metallicity-mass relation, to the extent that
it is now being used to gauge galaxy assembly in the intermediate-redshift
Universe \citep{KZ:99,CL:01}. Indeed, there is a general positive
correlation across the Hubble sequence of oxygen abundance with M$_B$,
that holds for over 10 orders of magnitude in M$_B$ \citep{HH:99}. It is not
as yet clear that a metallicity-luminosity relation holds for BCDs alone
\citep[Figure~7]{HH:99}. Similarly, our limited data do not support the
existence of a mass-metallicity relation for BCDs.

Since the seminal work of \citet{DS:86}, the extent to which mass
loss affects the evolution of low-mass galaxies has attracted much attention.
Such work is particularly interesting since low-mass systems are galactic
building blocks in hierarchical scenarios of galaxy formation, and could
contribute to the pollution of the intergalactic medium at early epochs.
Recent models for dwarf galaxies have addressed the role of dark
matter halos for the evolution of low-mass systems. \citet{MF:99}
find that mass loss occurs in all galaxies with gas masses below
$\approx 10^9$\,M$_\odot$, but complete loss of gas (blowaway) is possible
only for galaxies with very low HI masses, less than a few times
10$^6$\,M$_\odot$. The HI masses in the IT sample range from several times
10$^6$\,M$_\odot$, to several times 10$^9$\,M$_\odot$. They match extremely
well the HI mass range to which the models by \citeauthor{MF:99} apply.
This suggests that some mass loss occurs in all of the BCDs observed here. 
\citeauthor{MF:99} also point out that while the fraction of cool gas lost
in dwarf galaxies is generally small, most of the metals mixed with hot
gas are able to leave these galaxies. This helps to keep their metallicities
low. \citet{FT:00} re-iterate that even in the presence of
a metal-enriched outflow, the predominant mechanism for gas consumption is
the conversion of gas into stars, and not mass loss. 

Figure~\ref{zvsmhi} shows the relation between oxygen abundance and gas
(HI) mass for the IT sample of BCDs. Three dashed lines approximate the
models of \citet{FT:00} for dark-to-visible (HI) mass ratios of 0, 10, and
300, respectively. Our Figure may be compared with the equivalent Figure for
gas-rich Irregular galaxies in \citet[their Figure~4]{FT:00}.
The data for BCDs are generally found to overlap with the model parameters.
The BCD sample also shows considerable overlap with the ``nearby dwarfs" and
``low-surface-brightness dwarfs" samples to which \citeauthor{FT:00} compared
their models\footnote{A solar oxygen abundance of 12 + log(O/H) $\approx 8.9$,
consistent with that used by IT, is assumed for all three samples.}.
Our data thus lend additional support to the models of
\citeauthor{MF:99} and \citeauthor{FT:00}. This also means that most
galaxies classed as BCDs are dark-matter dominated.

\citeauthor{FT:00} predict total SFRs for a limited choice of parameters.
These SFRs are quite small, and allow a variation over a few orders of
magnitude, owing to differences in dark-to-visible mass ratios and time
evolution. The small global SFRs of BCDs that we measure here are broadly
consistent with these predictions, although our median SFRs appear to be
an order of magnitude higher than their model predictions, as is the fact
that we observe a wide range of SFRs. The effect of Malmquist bias on
our sample may partially explain this result, biasing our median SFRs to
higher values. \citeauthor{FT:00} emphasize that
there is a considerable spread in oxygen abundances for a given HI mass,
and that this can be understood because higher dark-to-visible mass ratios
cause a stronger gas compression, increase the SFRs, and in doing so, also
increase the oxygen abundances. A large difference in dark-to-visible mass
ratios among the BCDs can help us to reconcile the lack of correlation for
oxygen abundance versus HI mass and oxygen abundance versus W$_{20}$ while,
at the same time, SFR correlates with oxygen abundance and also shows a
general trend with HI mass, albeit with much scatter. 

\subsection{I~Zw~18 and SBS~0335$-$052}
 
I~Zw~18 and SBS~0335$-$052 once again stand out from the rest of the BCDs
in Figure~\ref{zvsmhi}. They actually fall in a ``forbidden" area of the
model parameter space. The fact that the two well-studied extremely metal
poor systems lie outside these models (i.e., in the unphysical region below
a dark-to-visible mass ratio of zero) reinforces their unusual properties.
This discrepancy may be explainable through uncertainties in the HI masses,
as, for I~Zw~18 and SBS~0335$-$052, these only need to
be lowered by 1--2 orders of magnitude to move them close to dark-matter
dominated terrain in this Figure. This may be feasible because both
I~Zw~18 and SBS~0335$-$052 have complex optical and radio morphologies
\citep{vZ:98,LV:80,Pus:01}. If the applicable HI mass is that of the
``main body", then we can lower the HI flux by a factor of 2 or so. A
smaller distance will also lower the HI mass (although in the case of
I~Zw~18 it could not be lowered by more than a factor of 2--3, since
otherwise it should be resolved more effectively into individual stars with
HST). Independent data on dark-to-visible mass ratios are also somewhat
uncertain due to the many assumptions involved. While in the case of
I~Zw~18, radio observations support a dark-to-visible mass ratio of about
10 \citep{vZ:98,LV:80}, the dynamical mass of SBS~0335$-$052 is only about
2--5 times that in gas and stars \citep{Pus:01}.
In any event, I~Zw~18 and SBS~0335$-$052, which are unlikely to be forming
their first stars now \citep{Sch:00} yet show oxygen abundances close
to the ``floor" of 12+log(O/H)$\approx$7.2 allowed by the models of
\citet{FT:00}, could be interpreted as galaxies that have a maximum
time-averaged mass-loss rate (since their masses are a few times
10$^8$\,M$_{\odot}$) and are thus unable to retain metals effectively.

Figure~4 of \citeauthor{FT:00} contains two galaxies, UGC~04483 and Leo~A,
with apparently similar properties to I~Zw~18 in terms of oxygen abundance,
HI mass, and dark-to-visible mass ratio. UGC~04483 is also present
in the BCD sample of IT used here. While \citeauthor{FT:00} show it in
the ``forbidden" region, the smaller distance derived in the meantime
from single-star photometry \citep{Dol:01}, has resulted in a downward
revision of its HI mass. This, combined with the 0.2~dex higher oxygen
abundance of IT, has moved UGC~04483 well into the dark-matter dominated
regime in our Figure~\ref{zvsmhi}. 

\citet{FT:00} present a case study of the SFH and the oxygen
enrichment of Leo~A. It will be interesting to see whether their model
also agrees with the broad parameters of I~Zw~18 (and SBS~0335$-$052).
One clear difference between Leo~A and I~Zw~18 (or SBS~0335$-$052), is
that the present-day SFR of the former is much smaller, by about 3 (4)
orders of magnitude, than the latter. 

It is worth recalling the location of I~Zw~18 and SBS~0335$-$052 on
Figure~\ref{zvssfr}. They appear to be too metal-poor for their SFR, or
too intensely star-forming for their oxygen abundance, compared to the
rest of the sample. Explaining this offset as an error in the measured
SFR is unlikely, requiring the actual SFR of I~Zw~18 (for example) to
be more than 7 orders of magnitude less than estimated here. The metallicity
of I~Zw~18 has been the subject of many observations and extensive
measurement, and explaining the offset as an error in the measured
metallicity is similarly unlikely. 

It will be interesting to see whether the models of \citeauthor{FT:00} can
reproduce the high SFRs, high gas-mass fractions, and low chemical
abundances of I~Zw~18 and SBS~0335$-$052. Considering the optical and
HI morphologies of I~Zw~18 and SBS~0335$-$052, it is possible that these
galaxies are involved in an interaction. The models of \citeauthor{FT:00}
do not apply to merger situations.

\section{Conclusions}

We investigated the SFRs of a small (50 objects) but well-defined sample
of BCDs, that of \citet{IT:99}. One advantage of studying this sample is
that ionized-gas abundances are readily available from the work of this team.
In addition, HI fluxes and profiles are in hand for many of the BCDs from
the work of Thuan and collaborators. This allows us to investigate
correlations between SFR, ionized gas chemistry, HI mass, and total mass.
The main results are summarized at the end of Section~\ref{results}. Here,
we draw some conclusions about some of the results that we think have
a wider implication.

We find that SFR$_{60\mu{\rm m}}$ agrees with SFR$_{\rm 1.4GHz}$ in BCDs.
This agreement exists over a wide range of SFRs (nearly five orders of
magnitude in the sample that we investigate), at low metallicity (on
average 0.1 of solar), and at small galaxy masses compared to those of
normal spirals. Unless several factors have conspired to yield
SFR$_{60\mu{\rm m}}$ = SFR$_{\rm 1.4GHz}$ in our sample, it is logical
to conclude that the calibration relations and their underlying assumptions
are valid for BCDs. Furthermore, the existence of this type of scaling
relation for a wide range of galaxies with different masses and sizes
(from spiral galaxies to late-type dwarfs), and with a wide range of
interactions (from merging to isolated), suggests that in terms of the
FIR/radio correlation, the local properties of individual star-forming
regions are more important than the global properties of the galaxies
that host them.

While BCD studies have benefitted greatly from recent advances in
far-infrared astronomy, and there is now no doubt that even very
metal-poor galaxies living in the present epoch contain dust, very little
is known as yet about their radio properties. Radio polarization observations
are non-existent but are needed to elucidate the nature of magnetic fields
in BCDs. Magnetic field orientations, therefore are not well known.
Because the FIR/radio correlation, valid for large spiral galaxies,
extends to BCDs, the presence of magnetic fields at some level seems
highly likely. The role of magnetic fields is still poorly understood,
however, and may contribute to the observed scatter in the FIR/radio
correlation.

Color-magnitude diagrams of nearby BCDs can serve as an external check on
the validity of the SFRs derived from FIR or radio luminosities. The
relevant data are difficult to obtain, since they require very high spatial
resolution, and while the few existing data indicate encouraging agreement,
additional CMD studies are required. Single-star photometry is, and in the
near future will remain, biased toward nearby objects with small SFRs due to
instrumental limitations. In Paper~II, we shall therefore investigate the
SFRs of BCDs in the IT sample using the H$\alpha$ luminosities as SFR
indicators.

The small scatter in SFRs about the mean FIR/radio ratio supports results
of CMD models that show that BCDs do not have highly episodic SFHs
\citep{Sch:01a}. Our results also indicate that the median SFRs of BCDs are
somewhat smaller than has been deemed typical in the past. The gas consumption
timescales are at a minimum a few hundred Myr, and more typically they are
of the order of a few Gyrs. In this respect, BCDs appear closer to death
than dwarf Irregulars \citep[compare, for example, with the recent depletion
time scales in][]{vZ:01}.

We find that there is no mass-metallicity relation among the BCDs in the
IT sample. The absence of a mass-metallicity relation for BCDs can be
understood with the help of differential winds \citep{Mar:94}.
The more recent models of \citet{MF:99} and \citet{FT:00}, which incorporate
dark matter and are applicable to the HI mass range that we observe in
BCDs, appear to predict well the global properties of BCDs, at least to
first order. An interesting prediction of these models is that all galaxies
with HI masses below 10$^9$\,M$_\odot$ are prone to lose mass through the
feedback of star formation. Interestingly, while there is not a lot of
cool gas lost, the loss of metals becomes more and more important the lower
the HI mass of the galaxy (with a peak in the metal-loss rate at an HI
mass of a few times 10$^8$\,M$_\odot$). This indicates a progressively
stronger violation of the assumption of a closed-box model for chemical
evolution toward low galaxy masses. We surmise that the absence of a
mass-metallicity relation in BCDs reflects just this metal-enriched
mass-loss. The basis for the mass-metallicity relation breaks down below
a gas mass of about 10$^9$\,M$_\odot$.

I~Zw~18 and SBS~0335$-$052 continue to stand out as extreme examples of
BCDs whenever their chemical properties are being investigated. In terms
of total SFR, HI mass, gas-consumption timescale and W$_{20}$, they fit in
well with other BCDs. But in terms of oxygen abundance versus SFR and HI
mass, they are extremes compared to the rest of the sample. I~Zw~18 and
SBS~0335$-$052 have gas masses of a few times 10$^8$\,M$_\odot$ where
dwarf galaxies, according to the models of \citeauthor{FT:00}, are
particularly vulnerable to the loss of heavy elements. A detailed prediction
of the SFH and the chemical enrichment of I~Zw~18 based on the work of
\citeauthor{FT:00} would be particularly interesting as a precursor to the
results of future deep single-star photometry. Chemo-dynamical models with
``gasping," rather than bursting, SFHs for I~Zw~18 are already in work by
\citet{Rec:02}. (``Gasping" in this sense means long episodes of
moderate star-forming activity separated by short quiescent intervals,
see e.g., \citeauthor{Tos:02} \citeyear{Tos:02}.)
It will be interesting to see what physical insights can
be gained from their simulations.

\section*{Acknowledgments}
AMH gratefully acknowledges support provided by NASA through Hubble
Fellowship grant HST-HF-01140.01-A awarded by the Space Telescope Science
Institute, which is operated by the Association of Universities for
Research in Astronomy, Inc., for NASA, under contract NAS 5-26555.

\begin{deluxetable}{lccclr}
\tablewidth{0pt}
\tablecaption{Properties of the BCD galaxy sample.
  \label{tab1}}
\tablehead{
\colhead{Name} & \colhead{RA} & \colhead{Dec} & \colhead{$S_{60\mu{\rm m}}$} & \colhead{$S_{\rm 1.4GHz}$} & \colhead{$v_{\odot}$} \\
 & \colhead{(J2000)} & \colhead{(J2000)} &\colhead{(Jy)} & \colhead{(mJy)} & \colhead{(km\,s$^{-1}$)} \\
 & & & & \colhead{NVSS (FIRST)} &
}
\startdata
UM 311    &  01 15 33.89  & $-00$ 51 31.9  &  2.50  & $17.2$ ($2.2$) &  1709  \\
UM 420    &  02 20 54.51  & $+00$ 33 23.6  &    & $<2.4$ ($1.1$) &  17514  \\
MRK 0600    &  02 51 04.57  & $+04$ 27 13.9  &    & $<2.6$ &  1008  \\
SBS 0335$-$052    &  03 37 44.04  & $-05$ 02 38.5  &    & $2.7$ &  4043  \\
MRK 1089    &  05 01 37.80  & $-04$ 15 27.7  &  4.06  & $31.7$ &  4068  \\
MRK 0005    &  06 42 15.53  & $+75$ 37 32.6  &  0.21  & $<2.8$ &  792  \\
MRK 0071    &  07 28 41.40 & $+69$ 11 26.0  &  3.30  & $19.0$ &  108  \\
MRK 1409    &  07 45 29.72  & $+53$ 25 37.6  &    & $2.5$ &  5576  \\
SBS 0749+568    &  07 53 41.53  & $+56$ 41 58.9  &    & $<2.3$ &  5471  \\
SBS 0749+582    &  07 53 50.93  & $+58$ 08 10.4  &    & $<2.4$ &  9548  \\
UGC 04483    &  08 37 03.00  & $+69$ 46 31.0  &    & $1.1$ &  178  \\
SBS 0907+543    &  09 11 08.51  & $+54$ 10 51.2  &    & $<2.5$ ($<0.8$) &  8124 \\
MRK 1416    &  09 20 56.19  & $+52$ 34 04.7  &    & $1.3$ ($<0.7$) &  2326  \\
SBS 0926+606    &  09 30 06.61  & $+60$ 26 51.8  &  0.27  & $2.7$ &  4122 \\
I Zw 018   &  09 34 01.92  & $+55$ 14 26.1  &    & $2.7$ &  742  \\
SBS 0940+544    &  09 44 16.72  & $+54$ 11 32.9  &    & $<2.3$ &  1638  \\
SBS 0943+561    &  09 46 46.59  & $+55$ 57 06.4  &    & $<2.3$ &  8850  \\
MRK 0022    &  09 49 30.45  & $+55$ 34 48.6  &    & $<2.2$ &  1551  \\
SBS 0948+532    &  09 51 31.81  & $+52$ 59 36.3  &    & $<2.3$ ($<0.9$) &  13880  \\
MRK 1434    &  10 34 10.15  & $+58$ 03 49.1  &    & $<4.0$ &  2269  \\
MRK 1271    &  10 56 09.00  & $+06$ 10 23.0  &    & $<3.2$ ($<1.0$) &  1013  \\
CG 0798    &  10 57 47.45  & $+36$ 15 39.3  &  0.53  & $7.3$ ($<1.0$) &  634 \\
MRK 0036    &  11 04 58.47  & $+29$ 08 22.1  &  0.23  & $2.4$ ($<0.9$) &  646 \\
MRK 0162    &  11 05 08.13  & $+44$ 44 50.1  &  1.29  & $16.0$ ($11.6$) &  6458  \\
SBS 1116+583B    &  11 19 24.79  & $+58$ 03 50.8  &    & $<2.1$ &  9905  \\
VII Zw 403   &  11 27 59.90  & $+78$ 59 39.0  &  0.38  & $1.4$ &  $-$100  \\
SBS 1128+573    &  11 31 16.33  & $+57$ 03 58.7  &    & $<2.3$ &  1787  \\
MRK 1450    &  11 38 35.62  & $+57$ 52 27.2  &  0.28  & $<2.0$ &  946  \\
UM 448    &  11 42 12.30  & $+00$ 20 03.0  &  4.14  & $33.5$ ($25.0$) &  5488 \\
MRK 0750    &  11 50 02.67  & $+15$ 01 23.8  &  0.44  & $4.3$ ($<1.0$) &  754  \\
UM 461    &  11 51 33.03  & $-02$ 22 22.8  &    & $<2.6$ &  1039  \\
UM 462    &  11 52 37.15  & $-02$ 28 08.5  &  0.94  & $3.0$ &  1199  \\
MRK 0193    &  11 55 28.35  & $+57$ 39 52.4  &    & $<2.7$ &  5157  \\
SBS 1159+545    &  12 02 02.37  & $+54$ 15 49.6  &    & $<2.3$ ($<1.0$) &  3537  \\
SBS 1205+557    &  12 08 28.18  & $+55$ 25 26.5  &    & $<2.5$ &  1740  \\
SBS 1211+540    &  12 14 02.50  & $+53$ 45 18.0  &    & $<2.9$ &  907  \\
SBS 1222+614    &  12 25 05.55  & $+61$ 09 08.5  &  0.29  & $3.6$ &  706  \\
MRK 0209    &  12 26 16.02  & $+48$ 29 36.6  &    & $4.5$ ($4.5$) &  281  \\
SBS 1249+493    &  12 51 52.37  & $+49$ 03 25.5  &    & $<2.2$ ($<0.9$) &  7296  \\
MRK 0059    &  12 59 00.35  & $+34$ 50 42.8  &  1.97  & $14.4$ ($6.0$) &  1089 \\
SBS 1319+579    &  13 21 22.47  & $+57$ 41 28.1  &    & $<2.9$ &  2060  \\
SBS 1331+493    &  13 33 22.93  & $+49$ 06 06.3  &    & $<2.4$ ($<0.9$) &  599 \\
MRK 1486    &  13 59 51.01  & $+57$ 26 24.1  &    & $1.9$ &  10136  \\
CG 0389    &  14 17 01.76  & $+43$ 30 13.4  &    & $4.3$ ($<1.0$) &  609  \\
CG 0413    &  14 22 38.84  & $+54$ 14 05.9  &    & $<2.3$ ($<1.0$) &  6176  \\
MRK 0475    &  14 39 05.43  & $+36$ 48 20.5  &    & $<2.7$ ($<1.0$) &  583  \\
CG 1258    &  14 44 00.15  & $+29$ 15 55.1  &    & $<2.2$ ($<1.0$) &  13605  \\
SBS 1533+574    &  15 34 13.31  & $+57$ 17 06.8  &  0.26  & $4.2$ &  3310 \\
MRK 0487    &  15 37 04.19  & $+55$ 15 48.3  &  0.30  & $<2.7$ &  665  \\
MRK 0930    &  23 31 58.29  & $+28$ 56 49.9  &  1.25  & $13.1$ &  5485  \\
\enddata
\end{deluxetable}

\begin{deluxetable}{lcclcc}
\tablewidth{0pt}
\tablecaption{Distances, SFRs and metallicities for the local BCD galaxies.
 \label{tab2}}
\tablehead{
\colhead{Name} & \colhead{D} & \colhead{$SFR_{60\mu{\rm m}}$} & \colhead{$SFR_{\rm 1.4GHz}$} & \colhead{12+log(O/H)} & \colhead{$M_{\rm HI}$} \\
 & \colhead{(Mpc)} & \colhead{(M$_\odot$yr$^{-1}$)} & \colhead{(M$_\odot$yr$^{-1}$)} & & \colhead{($10^8\,$M$_\odot$)} \\
 & & & \colhead{NVSS (FIRST)} &
}
\startdata
UM 311 & 21.3 & 1.5 & 1.1 (0.14) & $8.31\pm0.04$ & \\
UM 420 & 247.9 &   & $<$21.2 (9.23) & $7.93\pm0.05$ & \\
MRK 0600 & 14.5 &   & $<$0.077 & $7.83\pm0.01$ & $3.1\pm0.3$ \\
SBS 0335$-$052 & 58.3 &   & 1.3 & $7.29\pm0.01$ & $11.6\pm2.0$ \\
MRK 1089 & 59.8 & 18.8 & 16.2 & $8.04\pm0.06$ & \\
MRK 0005 & 15.5 & 0.065 & $<$0.095 & $8.04\pm0.04$ & $1.8\pm0.4$ \\
MRK 0071 & 3.9 & 0.065 & 0.041 & $7.83\pm0.02$ & $11.6\pm0.05$ \\
MRK 1409 & 83.0 &   & 2.5 & $8.01\pm0.04$ & $31.4\pm5.7$ \\
SBS 0749+568 & 81.6 &   & $<$2.2 & $7.85\pm0.05$ & $9.1\pm2.7$ \\
SBS 0749+582 & 139.6 &   & $<$6.8 & $8.13\pm0.03$ & \\
UGC 04483 & 3.2 &   & 0.0016 & $7.54\pm0.01$ & $0.075\pm0.01$ \\
SBS 0907+543 & 119.9 &   & $<$5.1 ($<$1.6) & $7.97\pm0.03$ & \\
MRK 1416 & 38.1 &   & 0.27 ($<$0.15) & $7.86\pm0.02$ & $6.8\pm1.1$\\
SBS 0926+606 & 63.4 & 1.4 & 1.6 & $7.91\pm0.01$ & \\
I Zw 018 & 15.0 &   & 0.087 & $7.18\pm0.01$ & $1.44\pm0.07$ \\
SBS 0940+544 & 28.7 &   & $<$0.27 & $7.43\pm0.01$ & \\
SBS 0943+561 & 130.4 &   & $<$5.6 & $7.74\pm0.06$ & \\
MRK 0022 & 27.5 &   & $<$0.24 & $8.00\pm0.01$ & \\
SBS 0948+532 & 202.2 &   & $<$13.4 ($<$5.4) & $8.00\pm0.01$ & \\
MRK 1434 & 38.1 &   & $<$0.84 & $7.79\pm0.01$ & \\
MRK 1271 & 11.9 &   & $<$0.064 ($<$0.021) & $7.99\pm0.01$ & \\
CG 0798 & 9.4 & 0.061 & 0.092 ($<$0.012) & $7.97\pm0.02$ & \\
MRK 0036 & 7.7 & 0.017 & 0.020 ($<$0.0079) & $7.81\pm0.02$ & $0.18\pm0.03$ \\
MRK 0162 & 96.6 & 15.5 & 21.4 (15.5) & $8.12\pm0.03$ & \\
SBS 1116+583B & 146.4 &   & $<$6.4 & $7.68\pm0.05$ & \\
VII Zw 403 & 4.4 & 0.0096 & 0.0040 & $7.69\pm0.01$ & $0.71\pm0.05$ \\
SBS 1128+573 & 32.1 &   & $<$0.34 & $7.75\pm0.03$ & \\
MRK 1450 & 19.7 & 0.14 & $<$0.11 & $7.98\pm0.01$ & \\
UM 448 & 78.5 & 32.9 & 29.5 (22.0) & $7.99\pm0.04$ & \\
MRK 0750 & 6.0 & 0.020 & 0.022 ($<$0.0049) & $8.11\pm0.02$ & $0.086\pm0.02$ \\
UM 461 & 10.2 &   & $<$0.039 & $7.78\pm0.03$ & \\
UM 462 & 17.2 & 0.36 & 0.13 & $7.95\pm0.01$ & \\
MRK 0193 & 79.4 &   & $<$2.4 & $7.81\pm0.01$ & $10.7\pm3.3$ \\
SBS 1159+545 & 56.8 &   & $<$1.1 ($<$0.47) & $7.49\pm0.01$ & \\
SBS 1205+557 & 32.0 &   & $<$0.36 & $7.75\pm0.03$ & $0.89\pm0.60$ \\
SBS 1211+540 & 20.1 &   & $<$0.17 & $7.64\pm0.01$ & $0.68\pm0.11$ \\
SBS 1222+614 & 16.5 & 0.10 & 0.14 & $7.95\pm0.01$ & \\
MRK 0209 & 5.8 &   & 0.022 (0.022) & $7.77\pm0.01$ & $0.79\pm0.05$ \\
SBS 1249+493 & 110.7 &   & $<$3.9 ($<$1.7) & $7.72\pm0.01$ & \\
MRK 0059 & 25.2 & 1.6 & 1.3 (0.55) & $7.99\pm0.01$ & \\
SBS 1319+579 & 36.7 &   & $<$0.56 & $8.10\pm0.01$ & \\
SBS 1331+493 & 15.1 &   & $<$0.078 ($<$0.030) & $7.82\pm0.05$ & $3.5\pm0.2$ \\
MRK 1486 & 150.6 &   & 6.2 & $7.88\pm0.01$ & \\
CG 0389 & 15.3 &   & 0.14 ($<$0.032) & $7.59\pm0.01$ & $2.7\pm0.2$ \\
CG 0413 & 94.5 &   & $<$3.0 ($<$1.2) & $7.75\pm0.01$ & \\
MRK 0475 & 13.6 &   & $<$0.072 ($<$0.027) & $7.93\pm0.01$ & $0.066\pm0.02$ \\
CG 1258 & 199.4 &   & $<$12.5 ($<$5.6) & $7.99\pm0.06$ & \\
SBS 1533+574 & 53.2 & 0.94 & 1.7 & $8.01\pm0.12$ & \\
MRK 0487 & 15.3 & 0.092 & $<$0.092 & $8.06\pm0.04$ & $0.80\pm0.14$ \\
MRK 0930 & 75.3 & 9.1 & 10.6 & $8.06\pm0.03$ & $30.2\pm0.3$ \\
\enddata
\end{deluxetable}

\begin{figure*}
\centerline{\includegraphics[width=15.0cm]{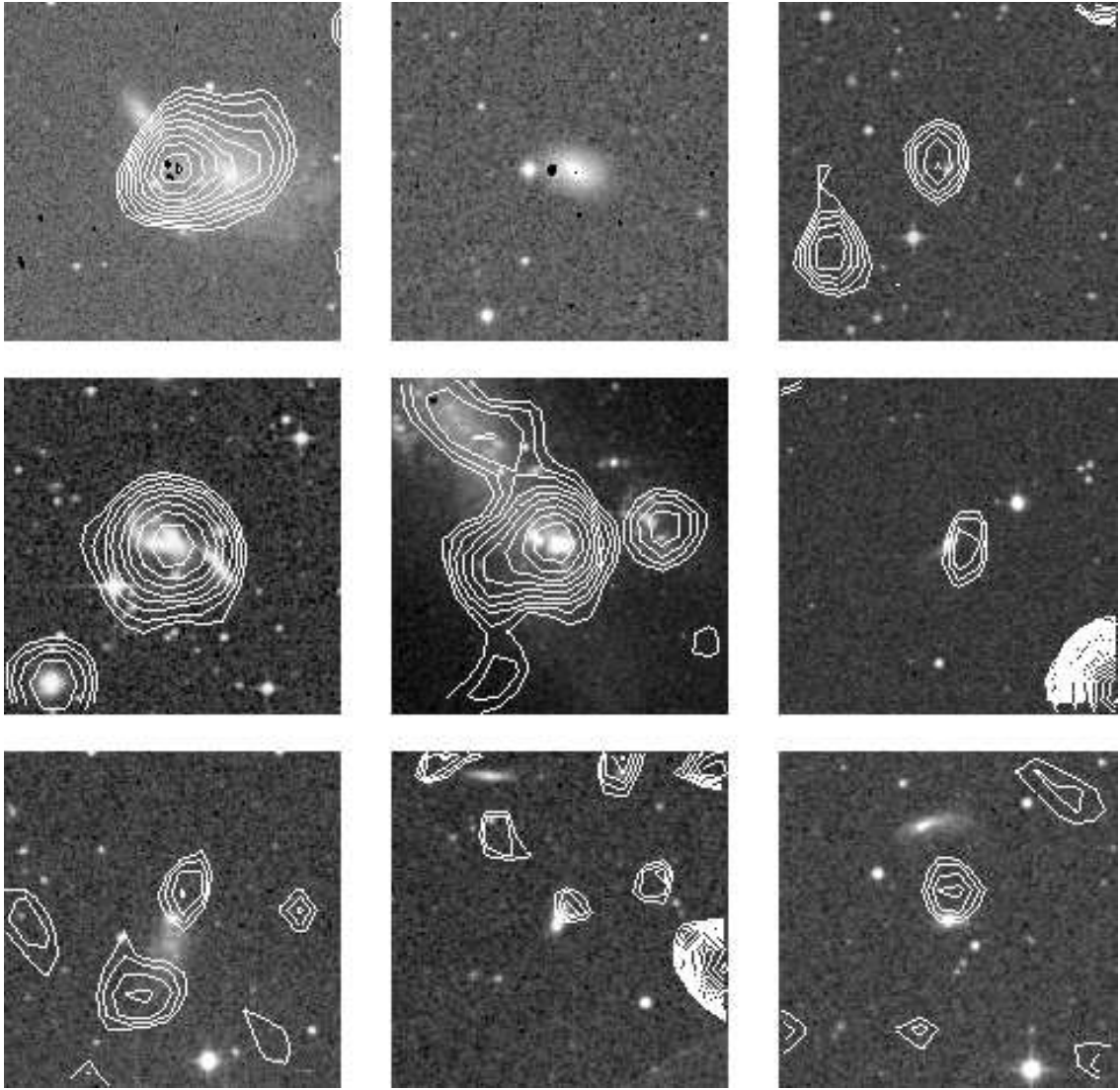}}
 \caption{Optical images from the Digitized Sky Survey (greyscale) with
1.4\,GHz radio contours in white from the NVSS overlayed.
The lowest contour level is 1\,mJy in all cases apart from
UGC~04483 (0.5\,mJy) and MRK~1416 (0.8\,mJy). For the 6 objects detected
at 1.4\,GHz by FIRST, contours are shown in black, with the lowest contour
level being 0.5\,mJy. Each image shown is about $4'\times4'$.
The 24 galaxies with NVSS or FIRST detections are shown in this Figure
(continued overleaf). From left to right, top to bottom the galaxies shown
above are:
UM~311; UM~420; SBS~0335$-$052; MRK~1089; MRK~0071; MRK~1409; UGC~04483;
MRK~1416; SBS~0926+606.
 \label{nvss1}}
\end{figure*}

\addtocounter{figure}{-1}
\begin{figure*}
\centerline{\includegraphics[width=15.0cm]{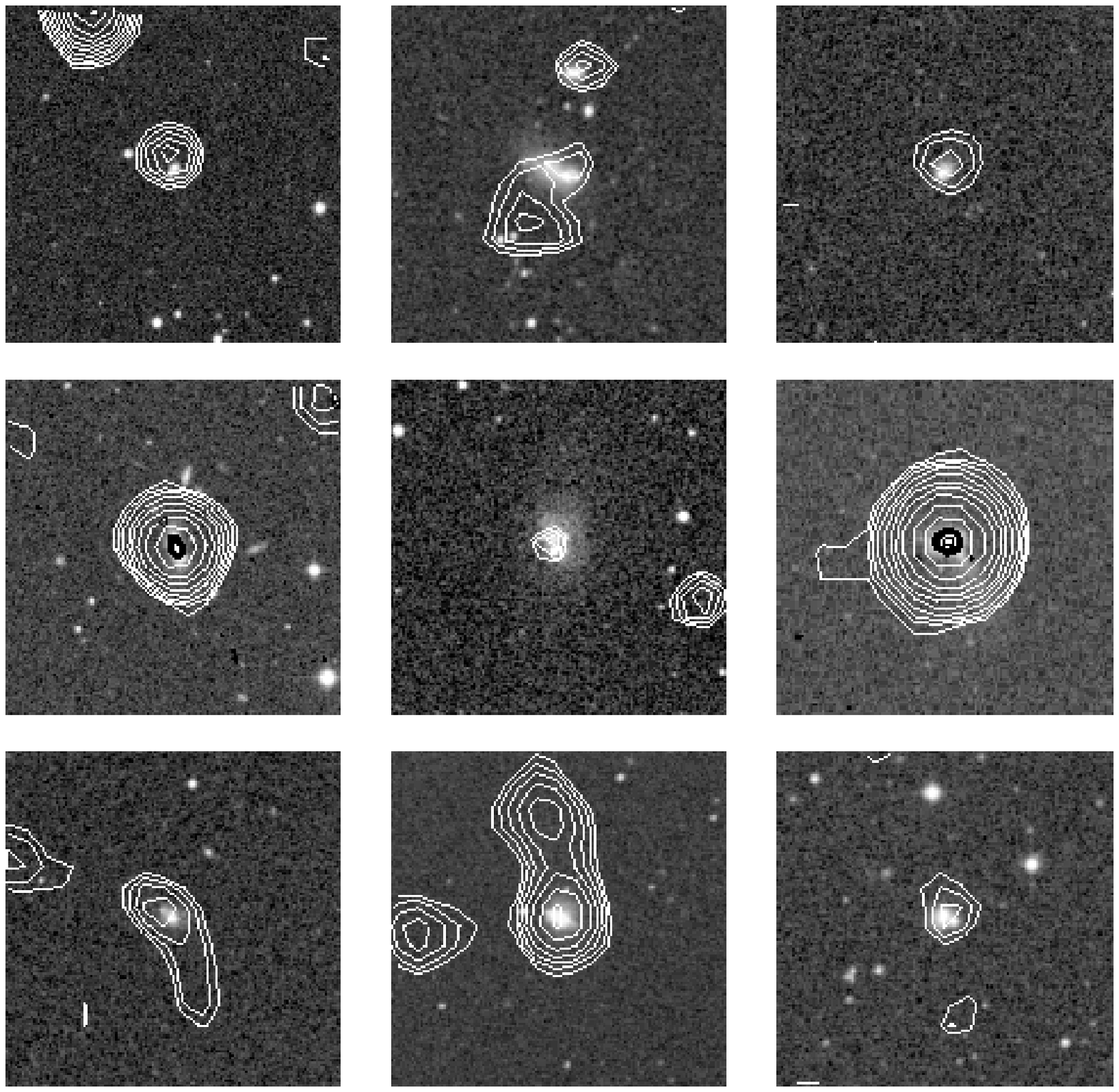}}
 \caption{Continued. From left to right, top to bottom the galaxies shown above
are: I~Zw~18; CG~0798; MRK~0036; MRK~0162; VII~Zw~403; UM~448; MRK~0750;
UM~462; SBS~1222+614.
 \label{nvss2}}
\end{figure*}

\addtocounter{figure}{-1}
\begin{figure*}
\centerline{\includegraphics[width=15.0cm]{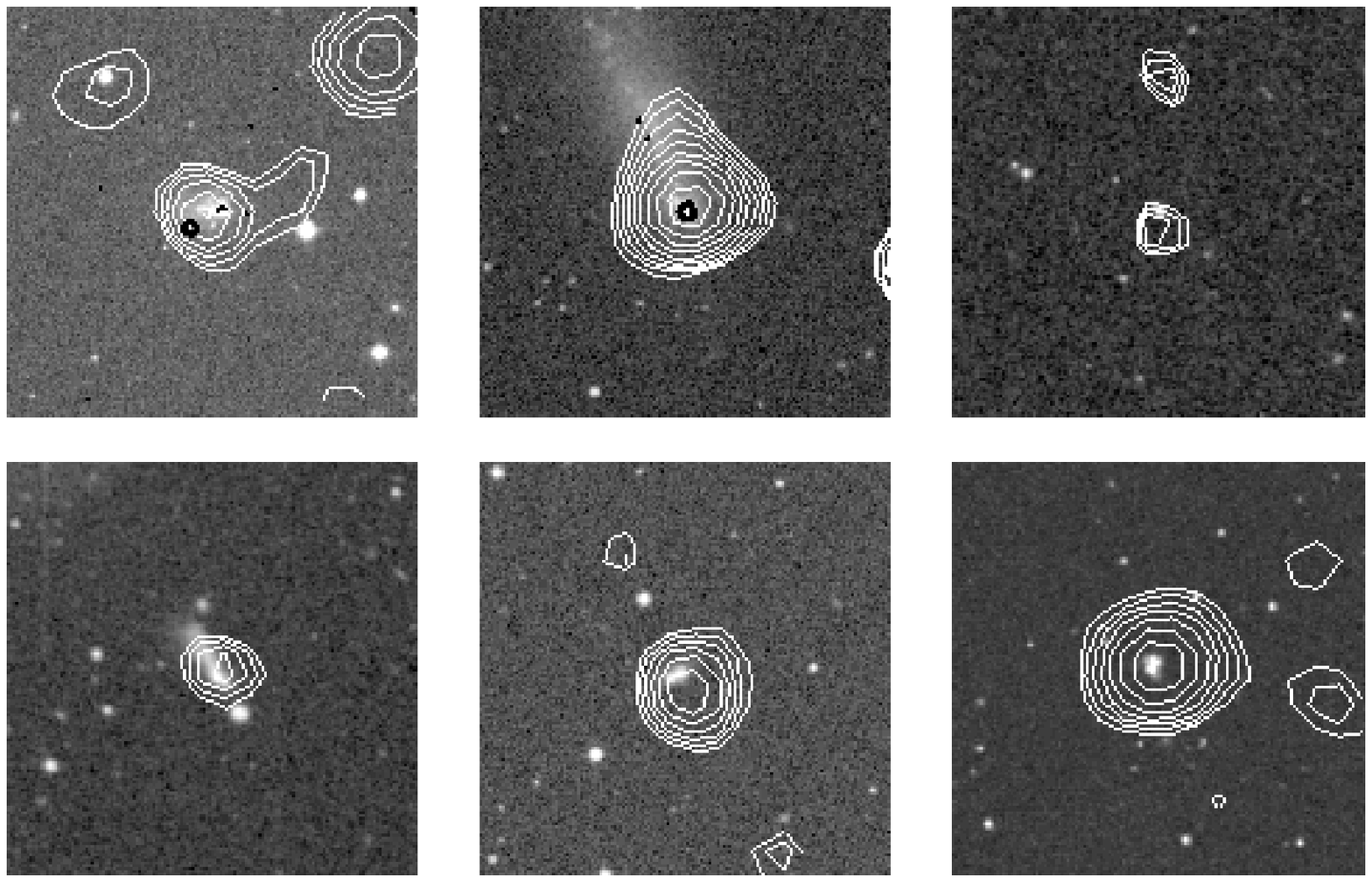}}
 \caption{Continued. From left to right, top to bottom the galaxies shown above
are: MRK~0209; MRK~0059; MRK~1486; CG~0389; SBS~1533+574; MRK~0930.
 \label{nvss3}}
\end{figure*}

\begin{figure*}
\centerline{\includegraphics[width=15.0cm]{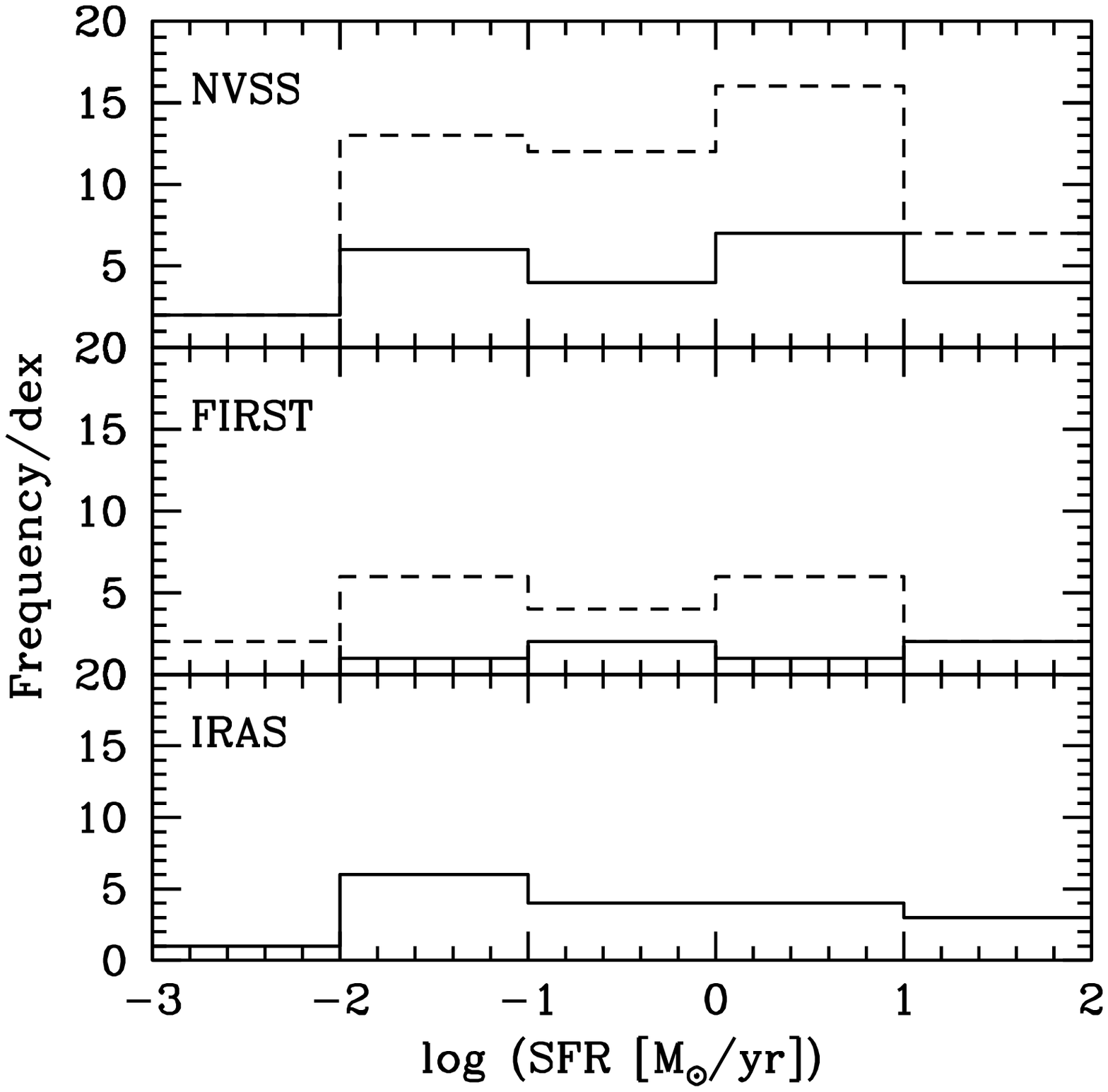}}
 \caption{Histograms showing the range of derived SFRs for the
sample. For the NVSS and FIRST (1.4\,GHz) estimates, the
inclusion of the upper limits as if they had been detections
are indicated by the dashed histograms. This figure primarily
indicates the broad range of SFRs exhibited in BCD galaxies.
 \label{sfrhist}}
\end{figure*}

\begin{figure*}
\centerline{\includegraphics[width=15.0cm]{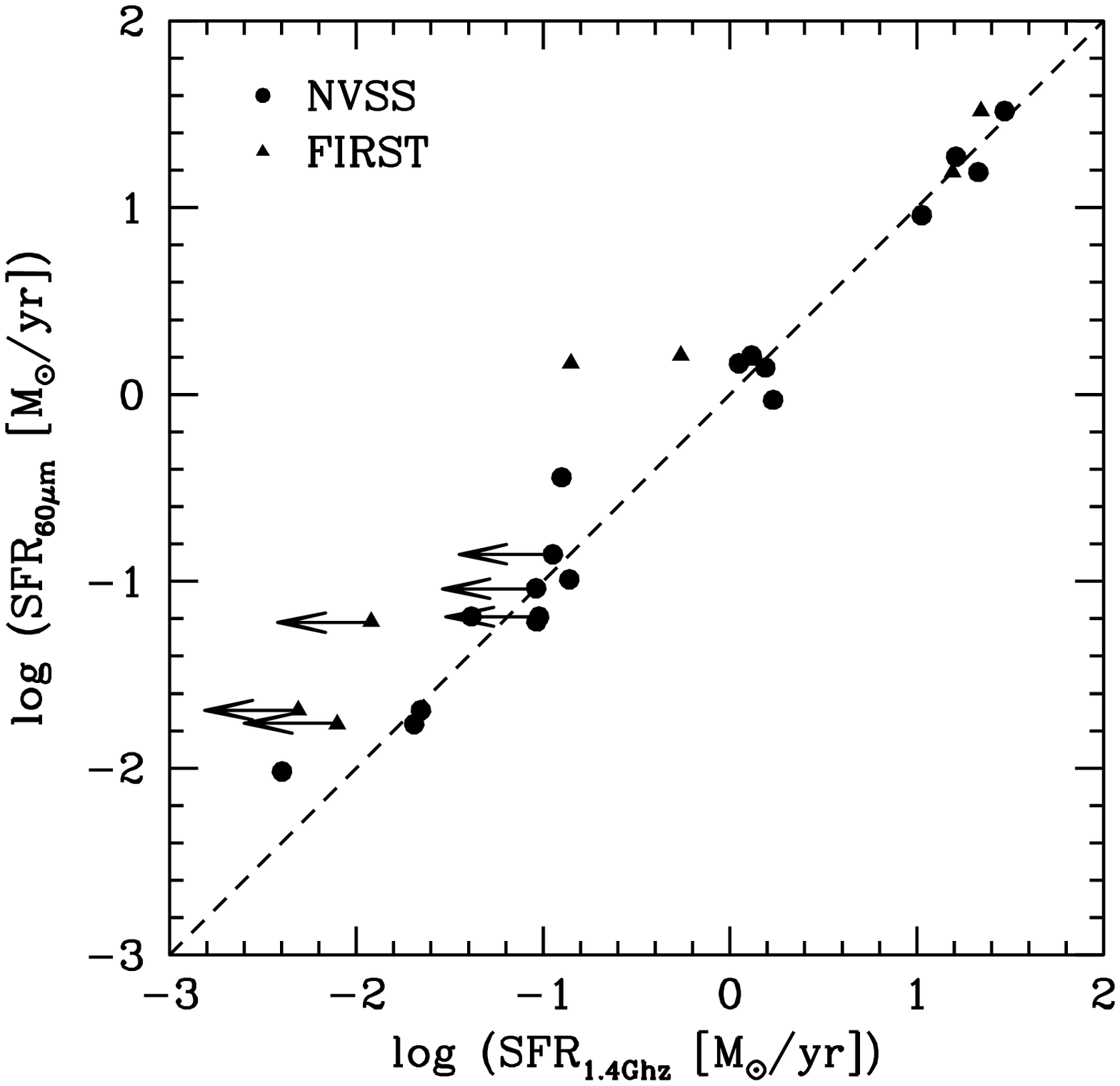}}
 \caption{Comparison of SFR$_{60\mu{\rm m}}$ with
SFR$_{\rm 1.4GHz}$ from both NVSS (circles) and FIRST (triangles).
The dashed line is the relation SFR$_{60\mu{\rm m}}$ = SFR$_{\rm 1.4GHz}$.
Note that the measurements of SFR$_{\rm 1.4GHz}$ derived from
FIRST are, in most cases, systematically lower than those from
NVSS. This is likely to be a result of NVSS having greater sensitivity
to the same extended structure as that detected by IRAS.
FIRST, although having much higher resolution than NVSS, detects
notably less of this more extended emission.
 \label{sfrvssfr}}
\end{figure*}

\begin{figure*}
\centerline{\includegraphics[width=15.0cm]{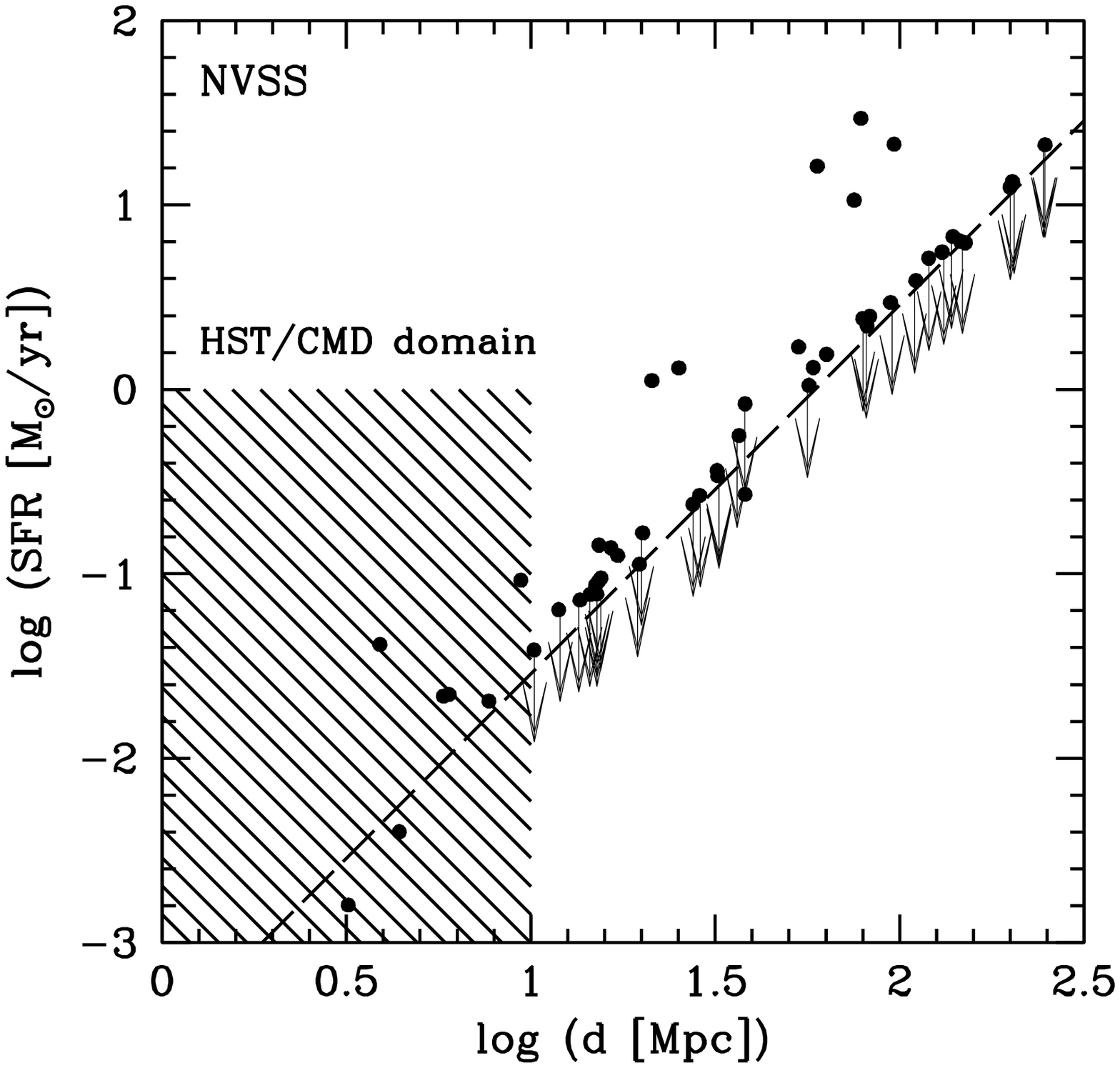}}
 \caption{SFR$_{\rm 1.4GHz}$ as a function of distance.
The dashed line corresponds to the sensitivity limit of the NVSS
catalogue (equivalent to a 2\,mJy detection). The hashed area
indicates the region (less than about 10 Mpc) within which it is
feasible to perform color-magnitude diagram analyses using
HST observations. Note the absence of any local BCDs with SFRs
greater than about 0.1\,M$_\odot$\,yr$^{-1}$.
 \label{sfrvsdist}}
\end{figure*}

\begin{figure*}
\centerline{\includegraphics[width=15.0cm]{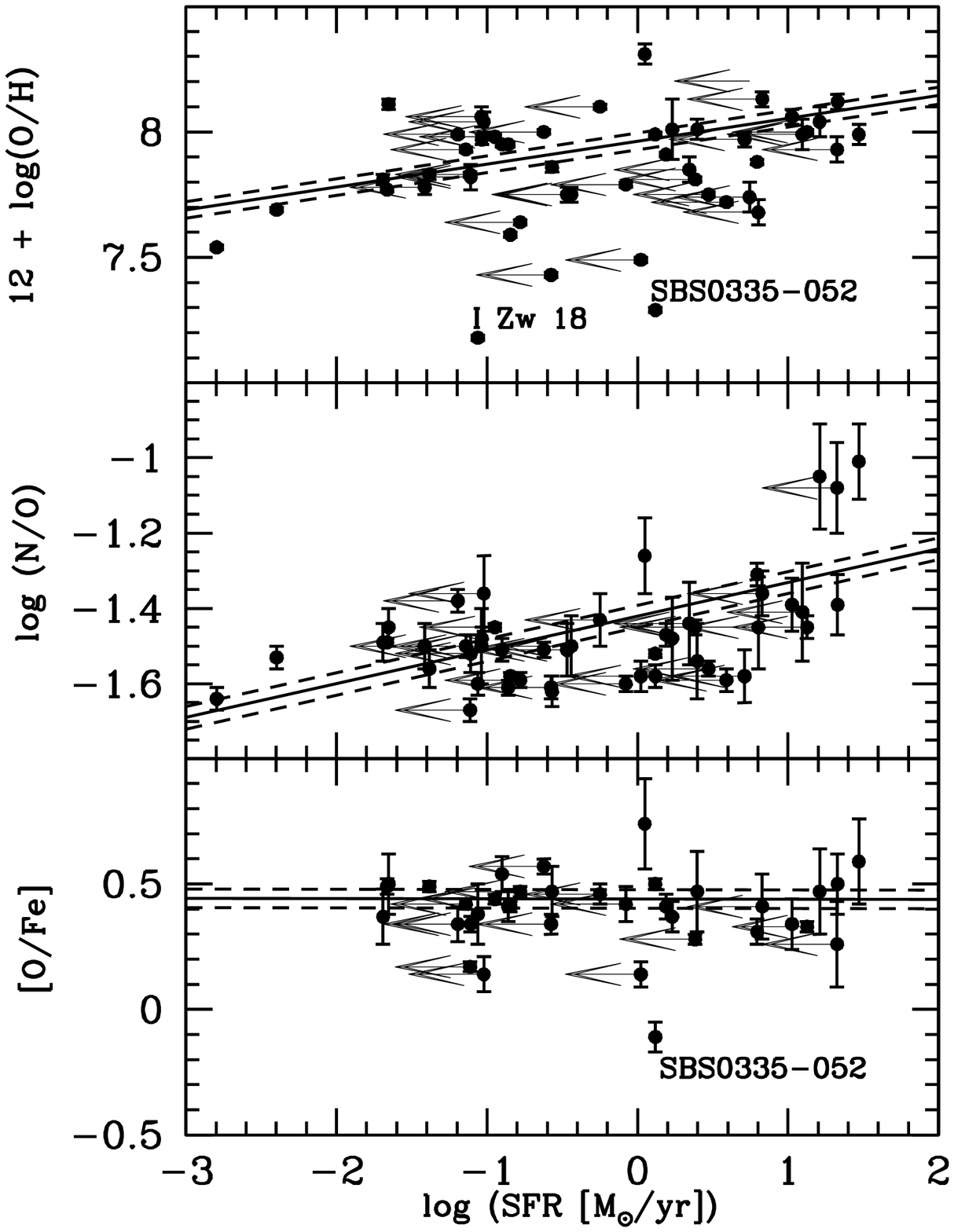}}
 \caption{Emission-line based abundance indicators for BCDs from IT
such as (12 + log(O/H)), log (N/O) and [O/Fe] as a function of SFR$_{\rm 1.4GHz}$.
A least squares fit to the objects with measurements (not upper limits)
of both SFR and abundance is shown, along with the $1\,\sigma$ uncertainties
for the intercept of the fit. Uncertainties on the slope are omitted
for clarity.
 \label{zvssfr}}
\end{figure*}

\begin{figure*}
\centerline{\includegraphics[width=15.0cm]{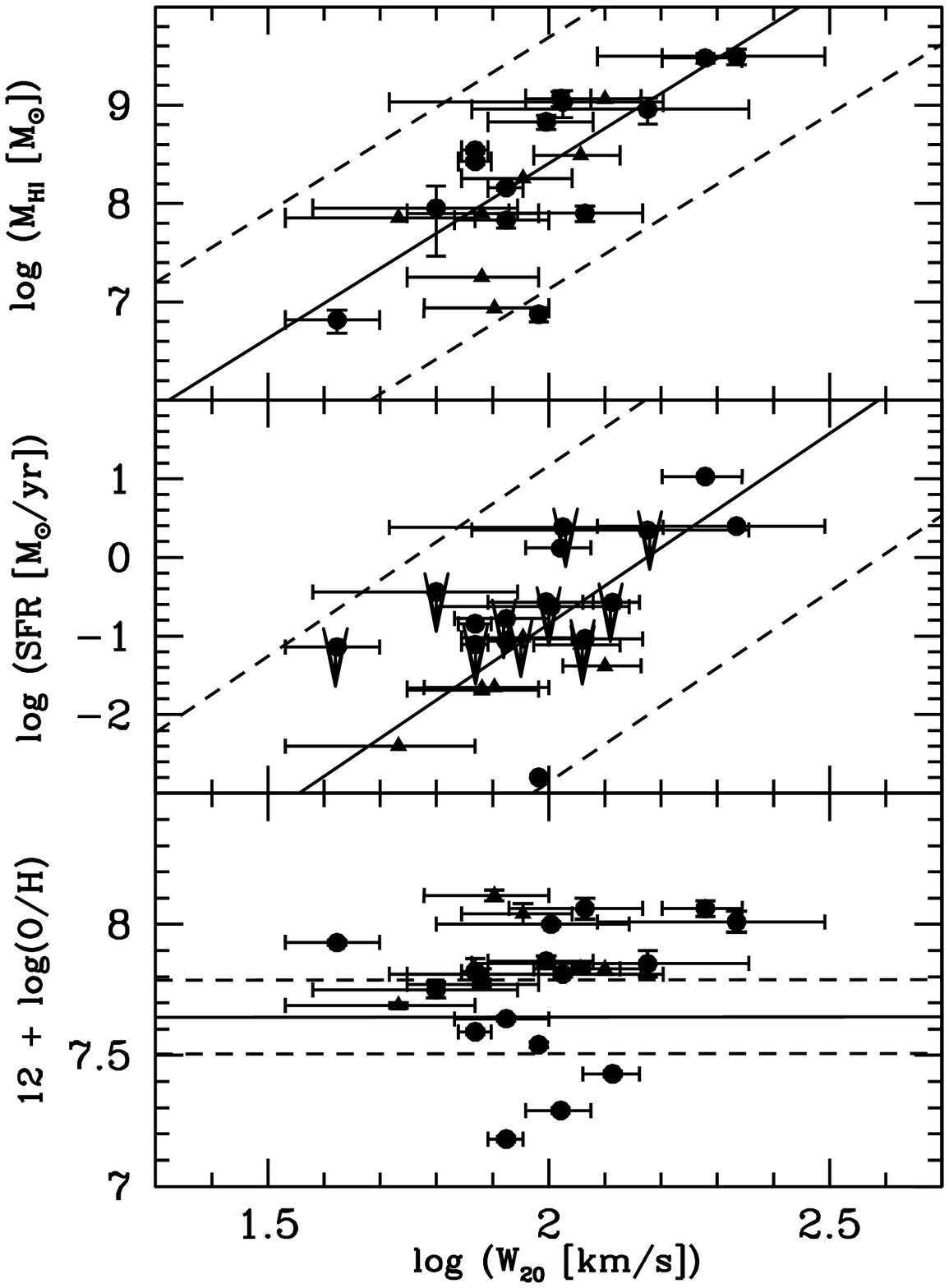}}
 \caption{SFR$_{\rm 1.4GHz}$ as a function of W$_{20}$, a rotational
velocity estimator based on HI emission line width. The circles refer 
to data from \protect\citep{Thu:99a} while the triangles us the data from 
\protect\citep{TM:81}. The least squares fits to the detections and
$1\,\sigma$ uncertainties for the intercept are again shown.
 \label{sfrvsw20}}
\end{figure*}

\begin{figure*}
\centerline{\includegraphics[width=15.0cm]{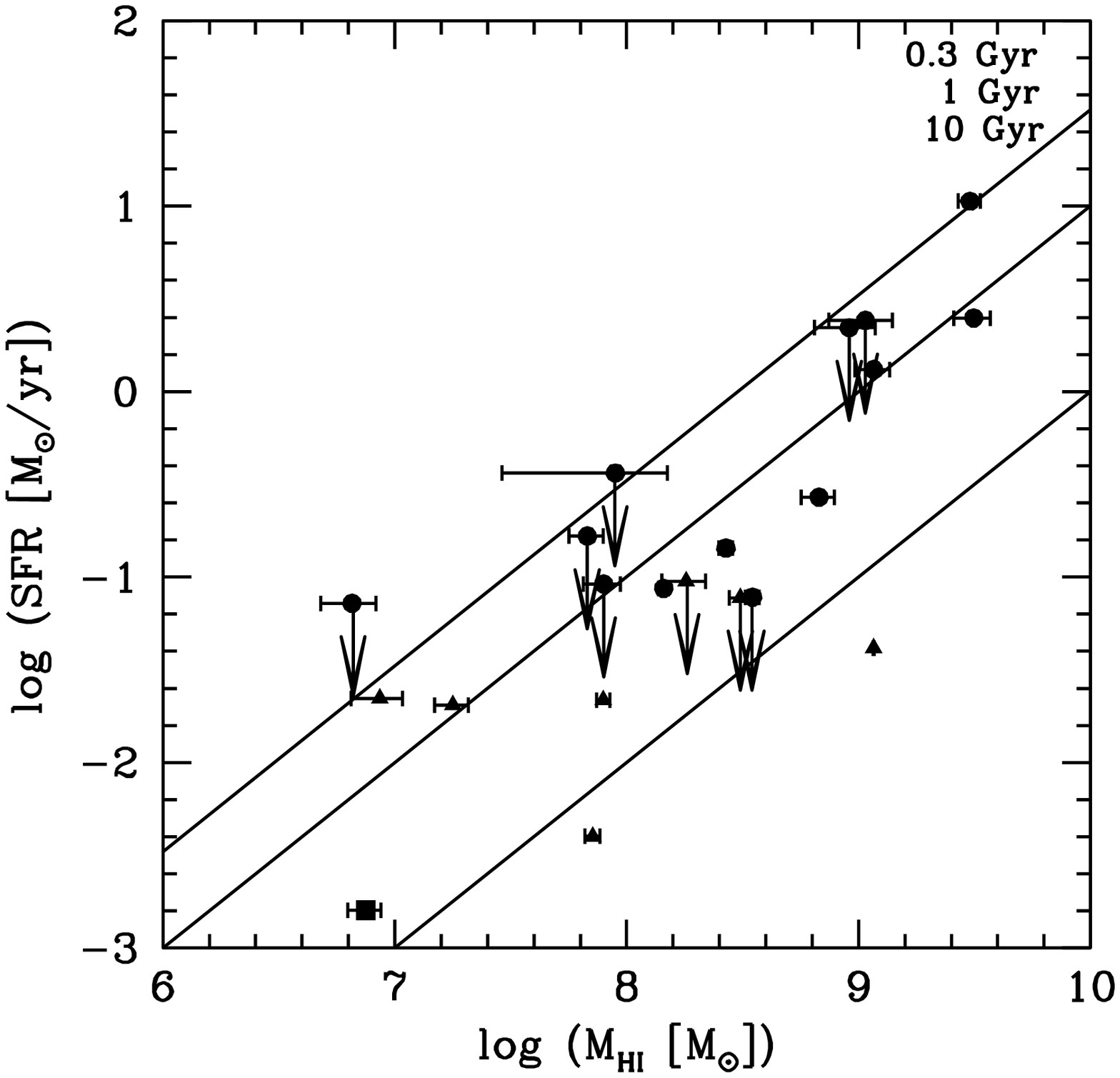}}
 \caption{SFR$_{\rm 1.4GHz}$ as a function of M$_{\rm HI}$
\protect\citep{Thu:99a,TM:81}. The square indicates data for the
metal-poor galaxy UGC~04483, using the HI flux and flux error
of Huchtmeier \& Richter (1986). Naive estimates of gas consumption time,
calculated simply by dividing the gas mass by the SFR, are shown.
 \label{sfrvsmhi}}
\end{figure*}

\begin{figure*}
\centerline{\includegraphics[width=15.0cm]{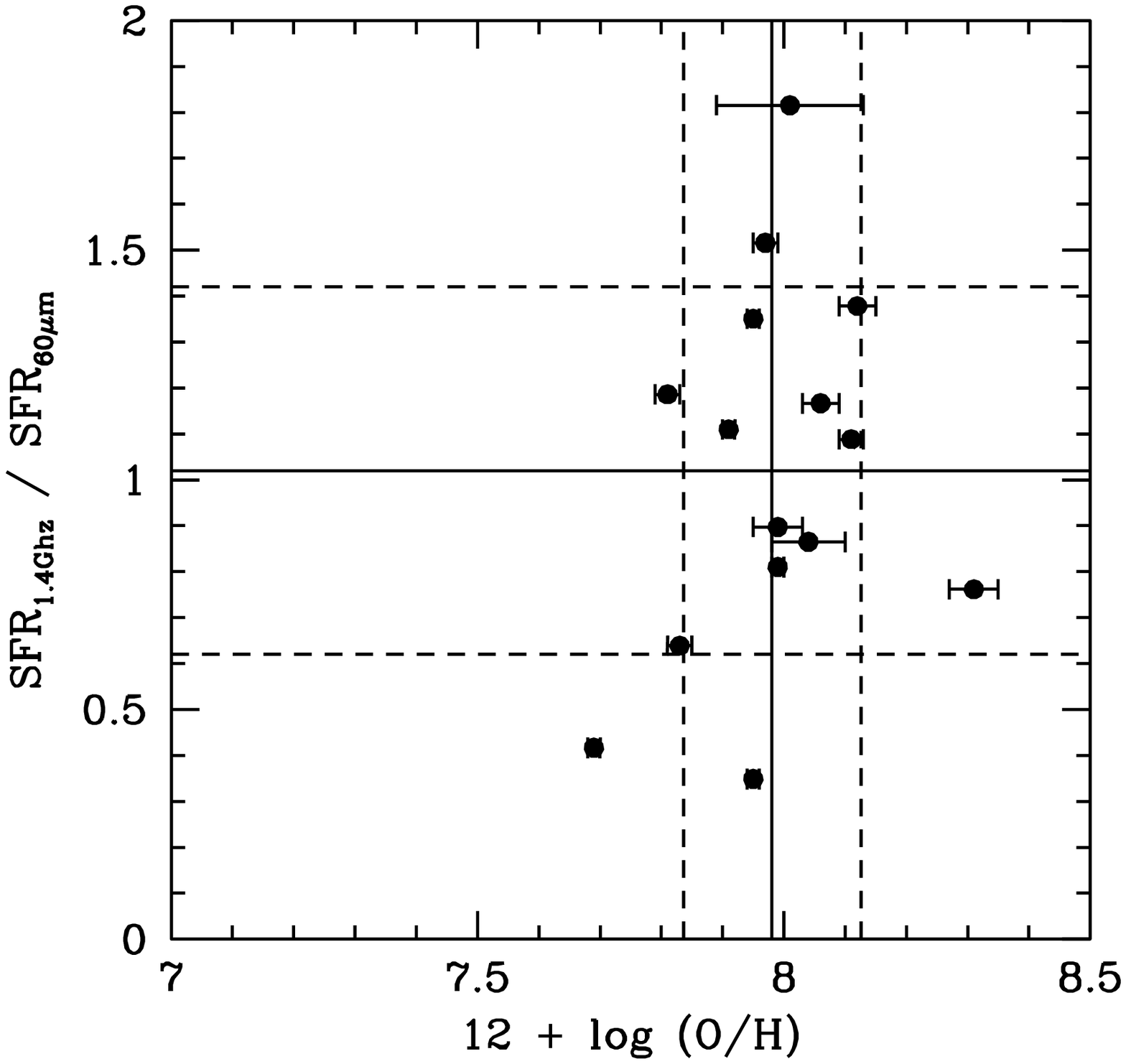}}
 \caption{To investigate any possible bias between the 1.4\,GHz
and 60\,$\mu$m SFR indicators, we explore the relationship 
between SFR$_{\rm 1.4GHz}$/SFR$_{60\mu{\rm m}}$ and
metallicity, 12 + log(O/H). The lines show the mean values (solid)
and the $\pm1\,\sigma$ values (dashed) for each of the quantities.
There does not seem to be any particular trend visible, supporting
the use and self-consistency of the two SFR indicators.
 \label{sfrratio}}
\end{figure*}

\begin{figure*}
\centerline{\includegraphics[width=15.0cm]{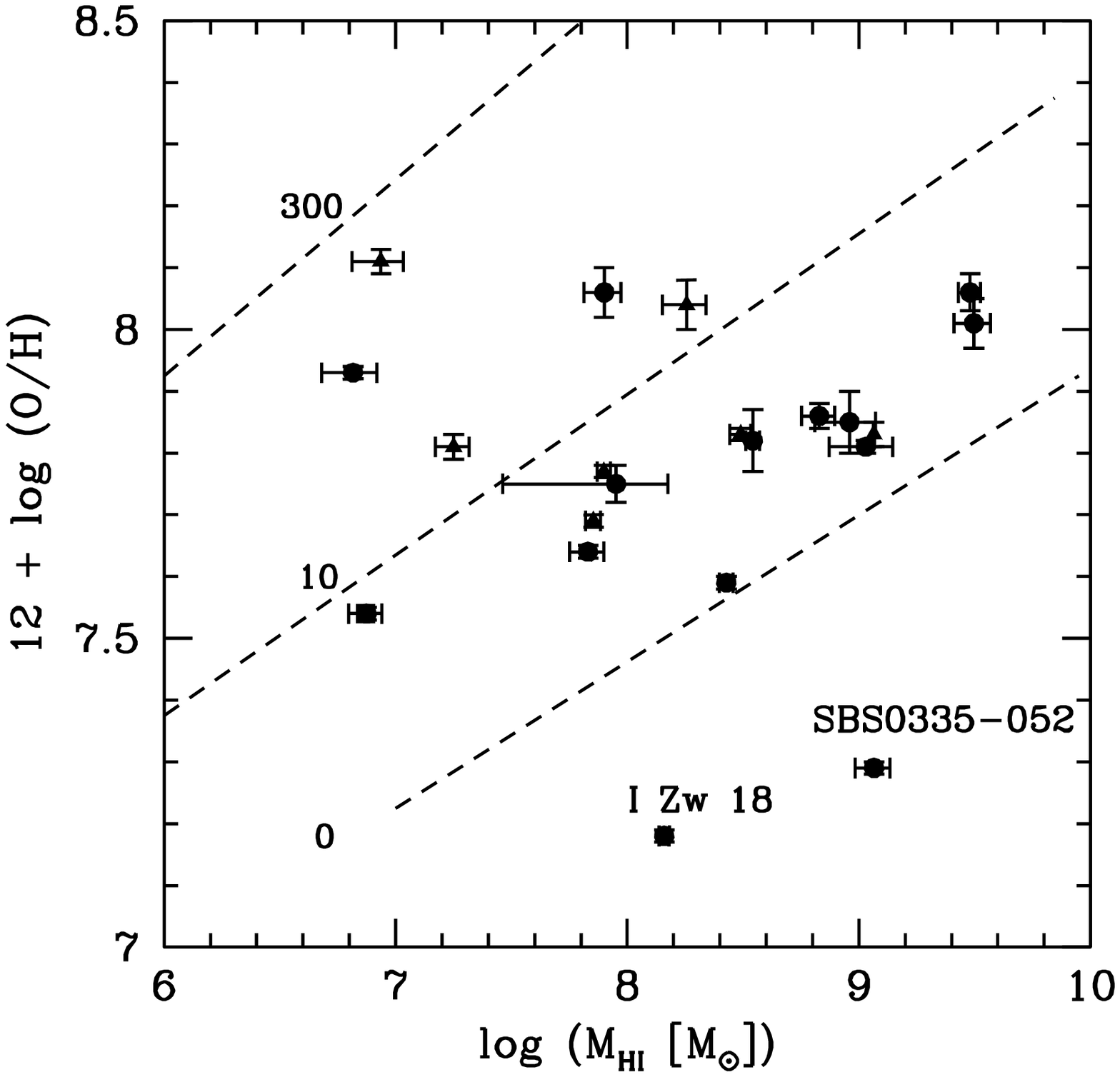}}
 \caption{Metallicity, 12 + log(O/H), as a function of M$_{\rm HI}$.
Dashed lines show dark-to-visible mass ratios after Ferrara \& Tolstoy (2000). 
Note the extremely metal poor galaxies I~Zw~18 and SBS~0335$-$052 again 
lying well below the rest of the population.
 \label{zvsmhi}}
\end{figure*}

\end{document}